\documentclass[letterpaper]{article}
\pdfoutput=1
\usepackage{jheppub}
\usepackage{amsmath}
\usepackage{graphicx}
\usepackage[config=altsf]{subfig}
\usepackage{feynmf}
\usepackage{dsfont}
\usepackage{array}
\usepackage{slashed}
\usepackage{afterpage}
\usepackage{appendix}
\usepackage{multirow}
\usepackage{wrapfig}

\newcommand{\eg}{{\it e.g.}}
\newcommand{\ie}{{\it i.e.}}

\usepackage[usenames,dvipsnames]{xcolor}

\usepackage{dcolumn}
\usepackage{slashed}

\newcommand{\be}{\begin{equation}}
\newcommand{\ee}{\end{equation}}
\newcommand{\bea}{\begin{eqnarray}}
\newcommand{\eea}{\end{eqnarray}}
\newcommand{\lsim}{\!\mathrel{\hbox{\rlap{\lower.55ex \hbox{$\sim$}} \kern-.34em \raise.4ex \hbox{$<$}}}}
\newcommand{\gsim}{\!\mathrel{\hbox{\rlap{\lower.55ex \hbox{$\sim$}} \kern-.34em \raise.4ex \hbox{$>$}}}}
\newcommand{\ltap}{\lsim}

\title{
\huge Top-philic $Z'$ forces at the LHC
}
\author[a]{Patrick J. Fox,}
\author[b,c,d]{Ian Low}
\author[a,c]{and Yue Zhang}
\affiliation[a]{Theoretical Physics Department, Fermilab, Batavia, IL 60510, USA}
\affiliation[b]{High Energy Physics Division, Argonne National Laboratory, Argonne, IL 60439, USA}
\affiliation[c]{Department of Physics and Astronomy, Northwestern University, Evanston, IL 60208, USA}
\affiliation[d]{Theoretical Physics Department, CERN, 1211 Geneva 23, Switzerland}

\abstract{
Despite extensive searches for an additional neutral massive gauge boson at the LHC, a $Z^\prime$ at the weak scale could still be present if its couplings to the first two generations of quarks are suppressed, in which case the production in hadron colliders relies on tree-level processes in association with heavy flavors or one-loop processes in association with a jet.  
We consider the low-energy effective theory of a top-philic $Z'$ and present possible UV completions.  We clarify theoretical subtleties in evaluating the production of a top-philic $Z'$ at the LHC and examine carefully the treatment of an anomalous $Z'$ current in the low-energy effective theory.  Recipes for properly computing the production rate in the $Z'+j$ channel are given. We discuss constraints from  colliders and low-energy probes of new physics. As an application, we apply these considerations to models that use a weak-scale $Z'$ to explain possible violations of lepton universality in $B$ meson decays, and show that the future running of a high luminosity LHC can potentially cover much of the remaining parameter space favored by this particular interpretation of the $B$ physics anomaly.}

\preprint{FERMILAB-PUB-18-005-T
\\ \rightline{NUHEP-TH/18-01}}

\begin{document}

\maketitle

\section{Introduction}

So far collider searches for new phenomena have not found any credible signals of new physics beyond the SM. These searches are extensive and broad, and have placed  significant constraints on many popular new physics models. In some cases exclusion limits at the LHC reached multi-TeV mass scales,  examples of which include the lower bound on a particular type of $Z^\prime$ at around 4 TeV \cite{Aaboud:2017buh} or a  limit on the mass of gluinos in supersymmetry at around 2 TeV \cite{Sirunyan:2017kqq}. Given that the initial gain in the reach of the LHC from the increased center-of-mass energy from 8 to 13 TeV is gradually tapering off beyond the first 10 fb$^{-1}$ or so,  the continuing absence of significant deviations in current data suggests we may be on a slow march toward any potential discoveries. Therefore, the time may be ripe to re-organize the search for  new physics and ponder over lingering opportunities -- Where is new physics at the LHC?

To answer the question, it is useful to recall that searches at hadron colliders, the LHC in particular, lean heavily on new physics having significant couplings to  partons inside the proton, and then subsequently decaying into high $p_T$ objects, whether visible or invisible, in the detectors. Thus, to find the new physics there are two obvious strategies to proceed. The first rests on the observation that the proton is mostly made of gluons and light-flavor quarks, therefore suppressing couplings to these partons would typically imply  a reduced production rate and/or open up new production channels that were previously subdominant. In particular, new channels often involve producing new physics in association with other SM particles. The second strategy makes use of the fact that, if new physics predominantly decays into soft objects, or outside of the detector, then the acceptance rate would diminish, rendering the signal more exotic and difficult to detect. These considerations motivated many proposals for  new physics at around or below the weak scale, such as those discussed in Refs.~\cite{deFlorian:2016spz,Alexander:2016aln}, and opened up new dimensions in searches for new physics. It then  becomes clear that, in spite of a tremendous amount of past and existing efforts, much remains to be done.

In this work we will consider one of the most extensively studied subjects in beyond-the-SM (BSM) landscape, the $Z^\prime$ gauge boson which appears in numerous extensions of the SM  \cite{Hewett:1988xc,Leike:1998wr,Langacker:2008yv}, and examine whether a light $Z^\prime$, in the order of a few hundreds GeV, could still be present at the LHC. As discussed already, one way to explain the absence of a $Z^\prime$ signal so far is to suppress, or turn off completely, its couplings to  the first two generations of quarks. Can we hide a light $Z^\prime$ at the LHC this way? 

There are a few reasons for the exercise we are doing. The first is to use this well-known object to demonstrate the need to think ``outside of the box" for BSM searches at this particular juncture in time. The goal is to evaluate current constraints on a $Z^\prime$ boson from a global perspective and devise future search strategies. In addition, we will see that the $Z^\prime$ boson considered in this work has novel collider signatures that are distinct from those previously considered in $Z^\prime$ searches. Secondly, the existence of a $Z^\prime$ boson is associated with an additional $U(1)$ gauge group. One of the most striking features of the SM is that all gauge anomalies  vanish. Adding additional gauge groups imposes constraints on the UV completion of the model such that the anomaly associated with the $Z^\prime$ must cancel. Usually this is brushed aside as an UV issue and irrelevant for the phenomenology in the low-energy,  as one can always introduce additional ``spectator fermions"  at very high energies whose sole purpose is to cancel the anomaly. On generic ground one does not expect these heavy spectator fermions to have an impact on the low-energy collider phenomenology of the $Z^\prime$. We will show that this is not always the case, especially when one considers a ``top-philic" $Z^\prime$ boson which couples to anomalous currents. Previous studies on collider phenomenology of a top-philic $Z^\prime$ \cite{Greiner:2014qna,Cox:2015afa,Kim:2016plm} contain some subtleties we aim to clarify in this work. 
The third reason is recently a top-philic $Z^\prime$ was invoked as a possible explanation to possible violations of the lepton universality in  $b\to s$ transitions \cite{Kamenik:2017tnu}. Needless to day, violations to lepton universality would  be a major discovery and it is important to evaluate its potential implications at the LHC.

Earlier discussions on an anomalous $Z'$ mostly focus on the couplings to massive electroweak gauge bosons in its decays \cite{Antoniadis:2009ze,Dudas:2009uq,Dudas:2013sia,Ekstedt:2017tbo}. More recent studies were done in the context of a light dark force mediator in low-energy probes \cite{Dror:2017ehi,Dror:2017nsg}, dark matter annihilations through a $Z'$ \cite{Zhang:2012da,Jackson:2013pjq,Ismail:2017ulg}, or electroweak gauge boson decays to $Z'$~\cite{Ismail:2017fgq}. Here we mainly focus on the collider phenomenology and discuss subtleties in calculating the $Z'$ production at the LHC, which is crucial to properly evaluating the experimental constraints and future search prospects in a high-energy collider~\footnote{These subtleties have also been addressed in \cite{2016PoaU}.}.  

This work is organized as follows.  We start the discussion in the next section, Section~\ref{sec:lowenergyEFT}, by working at energies where the only new state is a $Z'$ gauge boson and the theory is described by an EFT where the $Z'$ is coupled to the top quark and to SM leptons.  If the coupling of the $Z'$ to the top quark, or leptons, is chiral such a model is anomalous.  In Section~\ref{sec:twomodels} we consider in detail two possible UV completions of this low energy theory that fix the anomaly in different ways: an ``effective $Z'$" model and a ``gauge top model".  In Section~\ref{sec:LHCtopbounds} we investigate the LHC constraints on a top-philic $Z'$ from searches in multi-top final states, assuming all other couplings are negligible.  If the $Z'$ has couplings to SM leptons there are further constraints on the model which we discuss in Section~\ref{sec:pheno}.  Furthermore, these additional lepton couplings have the potential to explain recent anomalies seen in lepton universality violating observables in $B$ meson decays, without introducing any new flavor violation.  We show that the region of parameter in $Z'$ models that can explain the anomalies is smaller than previously thought.  In Section~\ref{sec:conclusions} we conclude.

\section{The low-energy effective theory}
\label{sec:lowenergyEFT}

The starting point of our discussion is the following effective Lagrangian, valid at the weak scale or below,  
\bea
\label{eq:eftdef}
\mathcal{L}_{eff} &=& \mathcal{L}_{\rm SM}  - \frac{1}{4} Z^\prime_{\mu\nu}Z^{\prime\,\mu\nu}
+ \frac{1}{2} M_{Z'}^2 Z'^\mu Z'_\mu  \nonumber \\
&&+ Z'_\mu\  \bar{t} \gamma^\mu (c_{t_L} P_L + c_{t_R} P_R) t+\sum_{i=e,\mu,\tau} Z'_\mu\  \bar{\ell}_i \gamma^\mu (c_{\ell_{i\,L}} P_L + c_{\ell_{i\,R}} P_R) \ell_i \ ,
\eea
where  $c_{\bullet_L}$, and $c_{\bullet_R}$ are free parameters at this stage. In the above we have used the notation $\bullet=\{t, \ell_e,\ell_\mu,\ell_\tau\}$. Also $P_{L,R}=(1\mp \gamma^5)/2$ are the usual projection operators. This is the effective Lagrangian for a top-philic $Z^\prime$ boson with couplings to leptons.

An effective  theory involving a $Z^\prime$ is usually thought of as descending from an abelian $U(1)^\prime$ gauge theory in the Higgs phase, where the gauge boson becomes massive. This could be achieved by introducing a complex scalar field whose vacuum expectation value $f$ breaks  $U(1)'$ spontaneously. The currents coupling to the $Z'$ are anomalous in general,  except for the carefully chosen parameters $c_{\bullet_L}=c_{\bullet_R}$. In the full theory the gauge invariance can be restored by introducing spectator fermions with the appropriate $U(1)'$ charges to cancel the anomaly. These spectator fermions could be heavy and integrated out of the  effective theory.  From the low-energy perspective,  non-conservation of the $U(1)'$ current is not necessarily a pathology, as the effective theory can still be consistently quantized as long as a cutoff is introduced \cite{Preskill:1990fr}.

These considerations justify using Eq.~(\ref{eq:eftdef}) as a ``simplified model" to study the collider phenomenology of a top-philic $Z'$ at the weak scale \cite{Greiner:2014qna,Cox:2015afa,Kim:2016plm}. There are some subtleties, however, that we wish to highlight. While a theory with an anomalous $U(1)'$ current can be a consistent effective theory, there is a particular class of diagrams involving the ``mixed-anomaly" between one $Z^\prime$ gauge boson and two gluons, which could have an important impact on the production mechanism of a top-philic $Z'$ at the LHC, because the $Z'$ does not couple to light flavor quarks and cannot be produced through the usual Drell-Yan process. Therefore, care must be taken when evaluating the contribution from the fermion triangle, and box, diagrams.

\begin{figure}[t]
   \centering
   \includegraphics[width=0.7\textwidth]{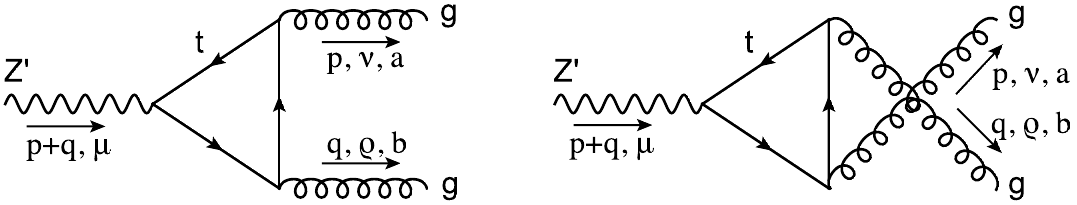} 
   \caption{Top quark contribution to the $U(1)'-SU(3)^2$ anomaly coefficient in the low energy theory of (\ref{eq:eftdef}).}
   \label{fig:triangle}
\end{figure}

Coupling of the $Z'$ boson with two gluons in the effective theory is induced at the one-loop level through the top quark triangle loop, as shown in the two diagrams in Fig.~\ref{fig:triangle}. In the effective theory the amplitude for each diagram is linearly divergent and a regulator as well as loop momentum routing scheme must be introduced for each to properly define the amplitude,
albeit the sum of the two diagrams is finite.
If the $Z'$ couples to an anomalous current, no regulator exists that would simultaneously preserve current conservation of all three external gauge bosons. Alternatively, the value of each diagram in Fig.~\ref{fig:triangle} depends on the routing of the loop momentum and, in an anomaly-free theory, the sum of the two diagrams does not depend on the momentum shift in the individual diagram. If the $U(1)'$ is anomalous, additional physical input is necessary to single out a particular momentum routing scheme. Two popular choices of scheme exist in the literature, which correspond to the consistent anomaly versus covariant anomaly \cite{Bardeen:1984pm}. In the context of our discussion, the consistent anomaly corresponds to symmetrizing with respect to all three external momenta, 
\be
-(p+q)^\mu {\cal M}_{\mu\nu\rho}^{ab}= p^\mu {\cal M}_{\nu\mu\rho}^{ab} = q^\mu {\cal M}^{ab}_{\nu\rho\mu} \neq 0 \ ,
\ee
which then implies the $SU(3)_C$ gauge invariance is lost. In the effective theory this is compensated by adding a Wess-Zumino term \cite{Wess:1971yu}
\be
\label{eq:wzw}
{\cal L}_{eff} \supset c_{WZ}\ g_X g_s^2\ \epsilon^{\mu\nu\rho\sigma} Z^\prime_\mu\left( G^a_\nu \partial_\rho G^a_\sigma +\frac13
  g_s \epsilon^{abc} G_\nu^a G_\rho^b G_\sigma^c\right)\ .
  \ee
The Wess-Zumino term gives a new contribution to the three-point amplitude such that the total amplitude, ${\cal M}_{\rm tot} = {\cal M}+{\cal M}_{WZ}$, satisfies $SU(3)_C$ gauge invariance
\be
\label{eq:gaugeinv}
p^\nu {\cal M}_{\mu\nu\rho}^{ab} = q^\rho {\cal M}^{ab}_{\mu\nu\rho} = 0 \ ,
\ee
which determines the coefficient $c_{WZ}$.
On the other hand, the covariant anomaly approach chooses a scheme that manifestly respects SM gauge invariance by maintaining Eq.~(\ref{eq:gaugeinv}) through a particular choice of momentum routing scheme in computing the triangle diagrams. In this case one can set $c_{WZ}=0$ in Eq.~(\ref{eq:wzw}). Both approaches lead to the same non-conservation of the $U(1)'$ current, in the limit that the fermion in the loop becomes massless,
\be
\label{eq:u1noncon}
-(p+q)^\mu {\cal M}^{ab}_{\mu\nu\rho} = {\rm Tr}(T^aT^b)\ (c_{T_L}-c_{T_R})\ \frac{ g_X g_s^{2}}{4\pi^2}\,\epsilon_{\nu\rho\lambda\sigma}\, p^\lambda q^\sigma\ ,
\ee
where we see explicitly that, when $c_{T_L}=c_{T_R}$, the $U(1)'$ current is vector-like and, therefore, conserved. 

In the low-energy effective theory, the appearance of a Wess-Zumino term can be interpreted as  arising from integrating out the heavy spectator fermions responsible for canceling the anomaly in the full theory. This is similar to integrating out the top quark in the SM and generating a Wess-Zumino term along the way \cite{DHoker:1984izu,DHoker:1984mif}. However, if one is not interested in the phenomenology of the spectator fermion, the distinction between the consistent versus covariant regularization in the simplified model is irrelevant, and both lead to the same amplitude in Eq.~(\ref{eq:u1noncon}) once the SM gauge invariance is imposed. The intricate interplay between the spectator fermion and the anomaly  will be demonstrated   later in this study.

One might argue that the calculation of the three-point amplitude in Fig.~\ref{fig:triangle} is irrelevant, as the Landau-Yang theorem forbids the coupling of an on-shell $Z'$ with two massless gauge bosons \cite{Landau:1948kw,Yang:1950rg}. There are two subtleties here that deserve to be clarified. First,  the selection rule arises from the inability to satisfy both the angular momentum conservation and the Bose symmetry: $J=1$ state of the on-shell $Z'$ requires the two massless gauge bosons to be in an anti-symmetric spin configuration while the Bose symmetry demands a symmetric total wave function.  For two gluons, however, there is a  twist in that the color degrees of freedom could provide an extra set of quantum numbers to symmetrize the wave function and satisfy Bose symmetry. But since the $Z'$ is a color singlet, the color indices of the two gluons must be in a symmetric combination and the Landau-Yang theorem is still valid.\footnote{A massive spin-1 boson that is color-octet can and will couple to two gluons on-shell~\cite{Keung:2008ve}.} The corollary of this discussion is that a top-philic $Z'$ cannot be singly produced on-shell from the gluon fusion at the LHC. Here lies the second subtlety: although the single production of an on-shell $Z'$ is forbidden, off-shell production  is still possible. In this channel, however, it is imperative to incorporate the width of the $Z'$ in a consistent fashion, for instance by adopting the complex mass scheme \cite{Denner:1999gp,Denner:2005fg,Denner:2014zga} and replacing $M_{Z'}^2 \to M_{Z'}^2 - i\Gamma_{Z'} M_{Z'}$ everywhere in the calculation \cite{Artoisenet:2013puc}. This has the effect of turning the production of an off-shell $Z'$ into a contact interaction between the gluons and the decay product of the $Z'$, which then becomes part of the QCD continuum background. It is an interesting question whether precision measurements of the relevant background could place meaningful constraints on the off-shell $Z'$ production, which is beyond the scope of the present work.

\begin{figure}[t]
\leftline{\hspace{0.7cm} \includegraphics[width=0.25\textwidth]{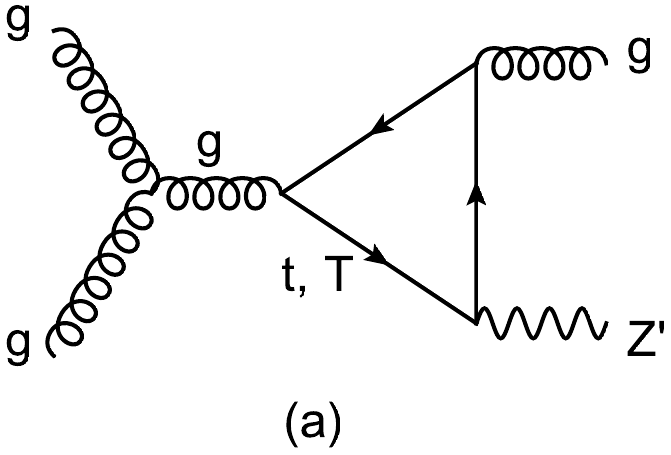} \hspace{1cm} \includegraphics[width=0.25\textwidth]{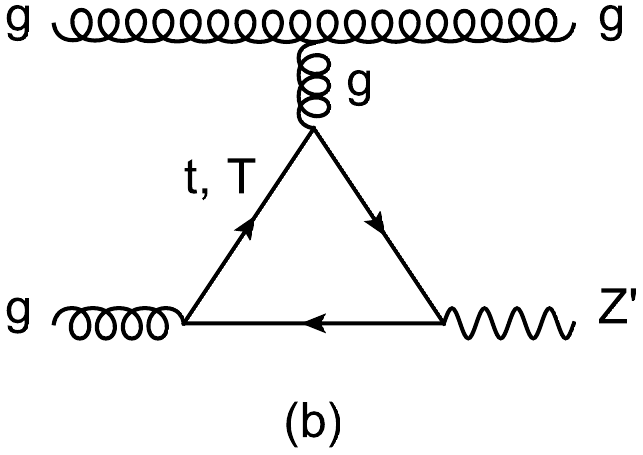} \hspace{1cm} \includegraphics[width=0.25\textwidth]{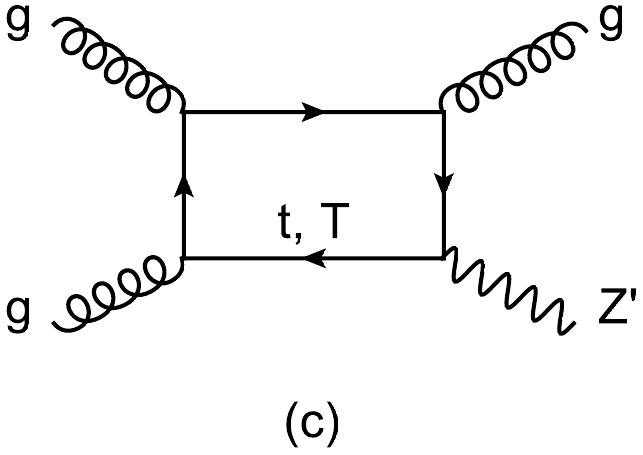}\vspace{0.6cm} }
\leftline{\hspace{0.7cm} \includegraphics[width=0.25\textwidth]{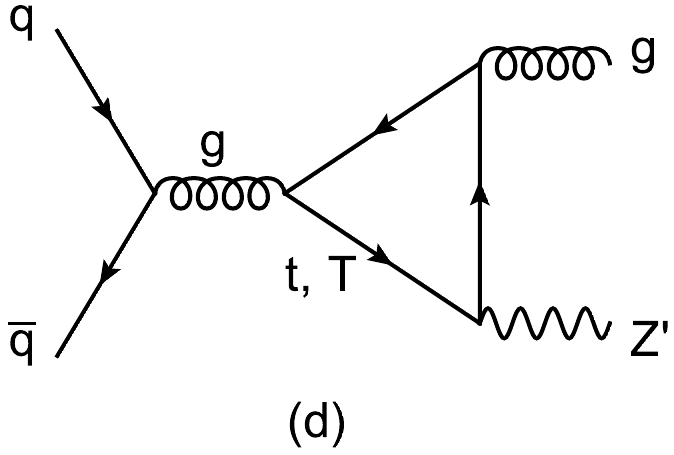} \hspace{1cm} \includegraphics[width=0.25\textwidth]{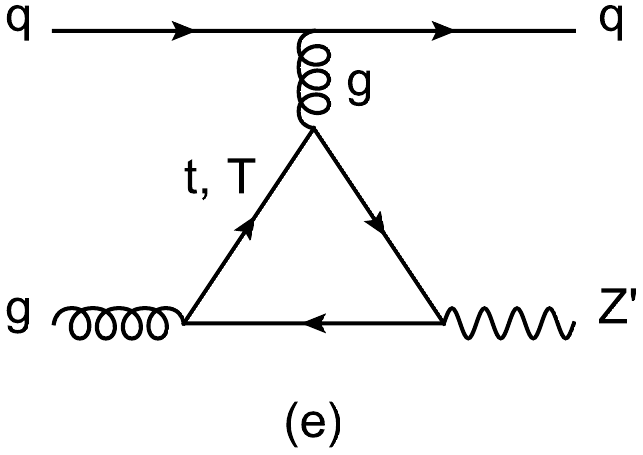}}
   \caption{\label{fig:feyn1}Feynman diagrams for loop-level $Z'$ production at the LHC in the effective theory.
\\ {\sf First row}: loop-level $g\bar g \to gZ'$ process;    {\sf Second row}: loop-level $q\bar q \to gZ'$ and $q g \to qZ'$ processes. Diagrams with inverted fermion arrows are not shown.}
   \label{fig:Feynman}
\end{figure}

In the end, to produce a top-philic $Z'$ on-shell we need to resort to associate production with other SM particles. Two production channels involving strong interactions are the tree-level $t\bar{t}Z'$ and the one-loop $Z'+j$ channels \cite{Greiner:2014qna,Cox:2015afa}, which will be considered in this work, although electroweak production such as the vector-boson fusion channel is also possible \cite{Antoniadis:2009ze}. The Feynman diagrams for the loop-induced production is shown in Fig.~\ref{fig:feyn1}. We see that the three-point coupling of one $Z'$ with two gluons features prominently, and must be dealt with carefully in the effective theory.

\section{Possible UV completions}\label{sec:twomodels}

In this section we consider possible UV completions to the effective Lagrangian in Eq.~(\ref{eq:eftdef}). The main purpose is to shed light on some of the subtleties in the calculation of the anomaly-induced couplings from the UV perspective.  It should be clear that, in addition to the SM, the minimal matter content must include a complex scalar $\Phi'$ charged under $U(1)'$ and neutral under $SU(2)_L\times U(1)_Y$, whose VEV  breaks $U(1)'$ spontaneously, as well as a vector-like pair of $SU(2)_L$ singlet fermions $(U'_L, U'_R)$ that carries $U(1)'$ charge. If the vector-like fermion mixes with the SM top quark, a low-energy theory like Eq.~(\ref{eq:eftdef}) can be obtained after integrating out the heavy fermionic mass eigenstate, which plays the role of the spectator fermion responsible for cancelling the $U(1)'$ anomaly. 

We consider two possible patterns for the mixing between  fermions with and without the $U(1)'$ charges. The first possibility is to introduce the mixing entirely through $U(1)'$ symmetry breaking effect, after the complex scalar $\Phi'$ gets a VEV, which is the  effective $Z'$ model presented in Ref.~\cite{Fox:2011qd}. In this scenario the coupling of the SM to the $Z'$ comes only through fermion mixing. In the second possibility, since the right-handed top quark is also an $SU(2)_L$ singlet, we could directly designate $U'_R$ as the right-handed top quark in the SM. This can be achieved by introducing an additional Higgs doublet $H'$ that is charged under $U(1)'$ to write down a gauge-invariant  Yukawa coupling between $Q_L$, the third generation left-handed doublet, and $U'_R$.  We call the second possibility the ``gauged-top" model, in which scenario there should be a second Higgs doublet $H$ that is not charged under $U(1)'$, to give mass to the remaining SM fermions.  

\begin{figure}[t]
\centerline{ \includegraphics[width=0.75\textwidth]{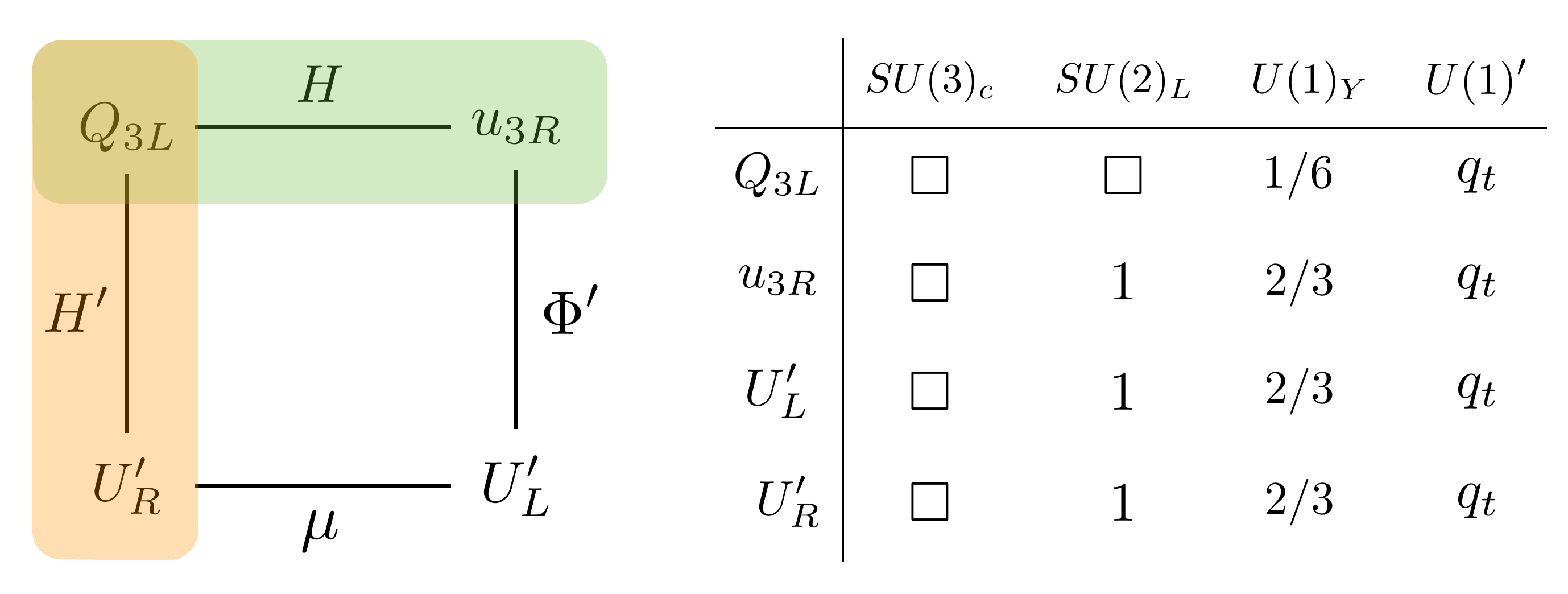}}
   \caption{Structure of interactions, and the charges of the relevant fields, in the general model defined Eq.~(\ref{Lint}).  The fermions on the top line of the diagram are uncharged under $U(1)'$ while those at the bottom are charged, the scalar $H$ has SM charges only, $\Phi'$ has $U(1)'$ charges only, and $H'$ has both.  The effective $Z'$ model, has the couplings in the vertical orange band turned off, while the gauged top model corresponds to those in the horizontal green band being zero.  Equivalently, the fields in the green band make up the SM-like states in the effective $Z'$ model and the fields in orange become the SM-like particles in the gauged top model. }
   \label{interactions}
\end{figure}

More specifically, we start with a $G_{\rm SM}\times U(1)'$ invariant theory with the following chiral fermion and scalar content.
\be
\begin{aligned}
 H' &:  (1, 2, -1/2, q_t), & \Phi' &:  (1,1,0, q_t),  & U'_{L} &: (3, 1, 2/3, q_t) , &  U'_{R} &: (3, 1, 2/3, q_t)\\
  H &: (1, 2, -1/2, 0), & Q_{3L} &: (3, 2, 1/6, 0) , & u_{3R} &: (3, 1, 2/3, 0)  \ ,
\end{aligned}
\ee
where $Q_{3L} \simeq (u_{3L}, d_{3L})$ and $u_{R3}$ are the third generation left-handed  quark doublet and the right-handed quark singlet in the SM, respectively, before electroweak symmetry breaking.  The lower indices $L,R$ indicates the chirality of fermions. In our notation, all the primed matter fields are charged under $U(1)'$, while the unprimed fields are neutral. The most general renormalizable interactions made of these fields takes the form
\be\label{Lint}
\begin{split}
\mathcal{L}_{int} =& \lambda_{H}\, \bar Q_{3L} \tilde H\, u_{3R} 
+ \lambda_{H'}\, \bar Q_{3L} \tilde H'\, U'_{R} + \lambda_{\Phi'}\, \bar U'_L u_{3R} \Phi' + \mu\, \bar U'_L U'_{R} + {\rm h.c.}  \\
&+g_X \left(\bar U'_L \gamma^\mu U'_L + \bar U'_{R} \gamma^\mu U'_{R} \right) Z'_\mu \ .
\end{split}
\ee
where for simplicity we suppressed the interaction of $Z'$ to the scalars. Without loss of generality, we set the charge $q_t=1$ hereafter. In the above $\lambda_{H'}$ gives the fermion mass mixing  after electroweak symmetry breaking, while $\lambda_{\Phi'}$ induces the mixing via $U(1)'$ breaking effect. Therefore the effective $Z'$ model corresponds to $\lambda_{H'}=0$ while the gauged-top model has $\lambda_{H}=0$.  Fig.~\ref{interactions} depicts the structure of fermion interactions via scalars in this model. 

When all the  couplings are non-zero, the physical right-handed top quark $t_R$ is a  linear combination of $u_{3R}$ and $U'_{R}$, with the orthogonal linear combination pairs up with $U'_L$ to form the vector-like top partner $(T_L, T_R)$ in the mass eigenbasis.  This is the $U(1)'$ anomaly-canceling spectator fermion. 
Notice that, since $H'$ is charged under both $U(1)'$ and the electroweak symmetry, its VEV  induces a tree-level mixing between the $Z'$ and the SM $Z$-boson, which are well constrained experimentally \cite{Cox:2015afa}. In the following subsections, we discuss the two limiting cases by varying the parameters in the above model which leads to the effective $Z'$ model and the gauged top model.

\subsection{Effective $Z'$ model}\label{effZ'}

The first limiting case is to set $\lambda_{H'} \to 0$. In this case, the $U(1)'$ and electroweak symmetry breaking are separate and there is no tree-level $Z-Z'$ mixing. The fermion mass matrix takes the form
\be
\mathcal{L}_{mass} = \begin{pmatrix} \bar{u}_{3L} & \bar U'_L \end{pmatrix}
\begin{pmatrix} 
\frac{1}{\sqrt{2}}\lambda_{H} v_H & 0 \\
\frac{1}{\sqrt{2}}\lambda_{\Phi'} v_{\Phi'} & \mu
\end{pmatrix}
\begin{pmatrix} u_{3R} \\ U'_R \end{pmatrix}
 \ ,
\ee
where the scalar field VEV's are $\langle H \rangle = v_H/\sqrt{2}\simeq 174$ GeV and $\langle \Phi'\rangle = v_{\Phi'}/\sqrt{2}$. After diagonalizing the mass matrix, there are two pairs of vector-like fermions denoted by $(t_L, t_R)$ and $(T_L, T_R)$, with physical masses $m_t$ and $M_T$. 
They are related to the fermion fields introduced in Eq.~(\ref{Lint}) via the rotation matrices,
\be\label{eq:rotation}
\begin{pmatrix} t_R \\ T_R \end{pmatrix} = \begin{pmatrix} \cos\theta_R & -\sin\theta_R \\ \sin\theta_R & \cos\theta_R \end{pmatrix} \begin{pmatrix} u_{3R} \\ U'_R \end{pmatrix}, \hspace{1cm}
\begin{pmatrix} t_L \\ T_L \end{pmatrix} = \begin{pmatrix} \cos\theta_L & -\sin\theta_L  \\ \sin\theta_L & \cos\theta_L \end{pmatrix} \begin{pmatrix} u_{3L} \\ U'_L \end{pmatrix} \ .
\ee
There are two rotational angles, $\theta_L$ and $\theta_R$, for the mixing between left- and right-handed fermion mixing, respectively.  They are related to each other and the physical parameters through,
\bea\label{eq:pararelations}
\begin{split}
\tan \theta_L &= \frac{m_t}{M_T} \tan \theta_R\,,\\ 
\lambda_{H} = \frac{1}{\sqrt{1+\frac{m_t^2}{M_T^2}\tan^2\theta_R}}\frac{\sqrt{2}m_t/v_H}{\cos\theta_R}\,,&\ \ \
\left|\lambda_{\Phi'}\right| = \frac{\sqrt2 g_X M_T}{M_{Z'}} \left(\frac{1-{m_t^2}/{M_T^2}}{\sqrt{1+\frac{m_t^2}{M_T^2}\tan^2\theta_R}}\right)\left|\sin\theta_R\right| \ .
\end{split}
\eea
where  the $Z'$ gauge boson mass $M_{Z'} = g_X v_{\Phi'}$.  Since the top Yukawa coupling, $\lambda_H$, is perturbed from its SM value the Higgs-gluon-gluon coupling will be altered from its SM value.  Global fits to Higgs properties, \eg\ \cite{Khachatryan:2016vau}, limit deviations in the top-Higgs coupling at 95\% C.L. to about 20\%, placing a bound of $|\sin\theta_R|\ltap 0.5$. Requiring that $|\lambda_\Phi|$ be perturbative up to 14 TeV places a constraint of $\lambda_\Phi(M_T\sim 1\,\mathrm{TeV}) \ltap 3$.  This will also limit the range of $\theta_{L,R}$ for given $M_{Z'}, M_T$ masses.

The $Z'$ couplings to the fermion mass eigenstates take the general form
\be\label{Lmaster}
\mathcal{L} = \bar{t} \slashed{Z}' (c_{t_L} P_L + c_{t_R} P_R) t + \bar{T} \slashed{Z}' (c_{T_L} P_L + c_{T_R} P_R) T  + \left(\bar{t} \slashed{Z}' (d_{L} P_L + d_{R} P_R) T + \mathrm{h.c.} \right)\ .
\ee
In this effective $Z'$ model, the above coefficients can be expressed using $\theta_R$ and the physical mass parameters,
\bea
\label{couplingseff}
\begin{split}
&c_{t_R} = g_X\sin^2\theta_R\ , \ \ \ c_{t_L} = g_X\sin^2 \left[\tan^{-1} \left( \frac{m_t}{M_T} \tan\theta_R \right)\right] \ , \\ 
& c_{T_R} = g_X\cos^2\theta_R \ , \ \ \  c_{T_L} = g_X\cos^2 \left[\tan^{-1} \left( \frac{m_t}{M_T} \tan\theta_R \right)\right] \ , \\ 
& d_R = -\frac{g_X}{2}\sin2\theta_R  \ , \ \ \ d_L = -\frac{g_X}{2}\sin \left[2\tan^{-1} \left( \frac{m_t}{M_T} \tan\theta_R \right)\right] \ .
\end{split}
\eea

\subsection{Gauged top model}\label{sec:gaugedtop}

In the second limiting case, we take $\lambda_{H} \to0$. The non-zero VEV's are
$\langle H'\rangle = v_{H}\sin\beta/\sqrt{2}$, $\langle H\rangle = v_{H}\cos\beta/\sqrt{2}$, and $\langle \Phi'\rangle = v_{\Phi'}/\sqrt{2}$. 
The VEV of $H'$ will induce a tree-level mixing between $Z$ and $Z'$. Their mass matrix takes the form
\begin{eqnarray}\label{eq:ZZ'mixing}
M_{ZZ'}^2 = \begin{pmatrix}
M_{Z, {\rm SM}}^2 & - g_X v_H \sin^2\beta M_{Z, {\rm SM}} \\
- g_X v_H \sin^2\beta M_{Z, {\rm SM}} & g_X^2 (v_H^2 \sin^2\beta + v_{\Phi'}^2)
\end{pmatrix} \ ,
\end{eqnarray}
where $M_{Z, {\rm SM}} = g v_H /(2 \cos\theta_W)$, $g$ is the $SU(2)_L$ gauge coupling and $\theta_W$ is the weak mixing angle.
The LEP experiment puts an upper limit on the shift of $Z$-boson mass as well as $Z-Z'$ mixing angle, which will be discussed in Section~\ref{sec:pheno}.

The SM-like right-handed top is now associated with $U'_R$ and the fermion mass matrix takes the form
\be
\mathcal{L}_{mass} = \begin{pmatrix} \bar u_{3L} & \bar U'_L \end{pmatrix}
\begin{pmatrix} 
\frac{1}{\sqrt{2}}\lambda_{H'} v_{H}\sin\beta & 0 \\
\mu & \frac{1}{\sqrt{2}}\lambda_{\Phi'} v_{\Phi'}
\end{pmatrix}
\begin{pmatrix} U'_R \\  u_{3R} \end{pmatrix}
 \ .
\ee
The rotations to the mass basis are similar to those of (\ref{eq:rotation}) but with $U'_R \leftrightarrow u_{3R}$.  Despite the relabelling of right-handed fields the structure of the mass matrix is the same as before.  Thus,
\bea\label{eq:pararelationsgaugedtop}
\begin{split}
\tan \theta_L &= \frac{m_t}{M_T} \tan \theta_R\,,\\ 
\lambda_{H'} = \frac{1}{\sqrt{1+\frac{m_t^2}{M_T^2}\tan^2\theta_R}}\frac{\sqrt{2}m_t/v_H}{\cos\theta_R\sin\beta}\,,&\ \ \
\left|\lambda_{\Phi'}\right| = \frac{\sqrt2 g_X M_T}{M_{Z'}} \sqrt{1+\frac{m_t^2}{M_T^2}\tan^2\theta_R} \cos\theta_R \ .
\end{split}
\eea
In the gauge top model, the coefficients defined in Eq.~(\ref{Lmaster}) take the following forms,
\bea
\label{couplingsg}
\begin{split}
&c_{t_R} = g_X\cos^2\theta_R\ , \ \ \ c_{t_L} = g_X\sin^2 \left[\tan^{-1} \left( \frac{m_t}{M_T} \tan\theta_R \right)\right] \ , \\ 
& c_{T_R} = g_X\sin^2\theta_R \ , \ \ \ 
c_{T_L} = g_X\cos^2 \left[\tan^{-1} \left( \frac{m_t}{M_T} \tan\theta_R \right)\right] \ , \\ 
& d_R = \frac{g_X}{2}\sin2\theta_R  \ , \ \ \ d_L = -\frac{g_X}{2}\sin \left[2\tan^{-1} \left( \frac{m_t}{M_T} \tan\theta_R \right)\right] \ .
\end{split}
\eea
Here the right-handed couplings are different from those in the effective $Z'$ model, Eq.~(\ref{couplingseff}), 
while the left-handed couplings remain the same.  This is because the SM top quark-like fermion is directly charged under the $U(1)'$.
Naively, this seems to indicate that the $Z'$ is more strongly coupled to the SM top quark in this model. 
However, the LEP experiment sets a strong constraint on the $Z-Z'$ mixing and in turn the gauge coupling $g_X$ (see Section~\ref{ZZpMix} and Fig.~\ref{fig:combined}) if the $Z'$ mass is close to the weak scale. 

\section{Top-philic $Z'$ production channels at LHC}\label{sec:Z'prod}

Having discussed the possible UV completions of a top-philic $Z'$, we now explore in this section its production at the LHC. The relevant interaction terms that determine the production rate of the $Z'$ are those of Eq.~(\ref{Lmaster}).  
The dominant production channels of such a top-philic $Z'$ boson at the LHC include
\begin{itemize}
\item  The tree-level process, $pp\to t\bar t Z'$ which depends only on $c_{t_L}, c_{t_R}$ couplings, and is dominated by gluon initiated states. We neglect the subdominant processes such as $pp\to t WZ'$ and $pp\to t j Z'$.
\item  Loop-level processes, $gg\to g Z'$, $q\bar q \to gZ'$, $q g \to qZ'$ and $\bar q g \to \bar q Z'$. The fermions $t$ and $T$ both contribute to the loop level processes, as can be seen in Fig.~\ref{fig:Feynman}.  As we will see below, when working in the low energy theory below the $T$ mass one must be careful to correctly include non-decoupling effects of $T$ running in the loop.
\end{itemize}

For numerical determination of these production cross sections we create the $Z'$ model file using {\tt FeynRules}~\cite{Christensen:2008py}, and compute the cross section for tree-level process $pp\to t\bar t Z'$ using {\tt MadGraph5}~\cite{Alwall:2011uj}. For calculating the loop processes, we resort to {\tt FeynArts/FormCalc}~\cite{Hahn:1998yk} for calculating the parton level cross sections, and then use  NNPDF~\cite{Ball:2014uwa, Carli:2010rw} for calculating the LHC production cross section at 13 TeV.

As will be shown in this section and the next two, the cross section for the loop production of $Z'+jet$ could be comparable to that of the tree level $Z'+t\bar t$ production, and it has a strong impact on the $Z'$ search at the LHC.

\subsection{Effect of heavy $T$ on loop-level $Z'$ production}\label{sec2.1}

We now elaborate on the effect of heavy fermion $T$ on the loop production channels of $Z'$, and discuss the consistent way to decouple $T$ as its mass becomes large, see also~\cite{2016PoaU}. As one would expect, the cancellation of gauge anomalies play an important role in this procedure.
The $Z'$ can couple differently to left- and right-handed quarks.  However, if we start from a non-anomalous UV theory and keep the heavy fermion $T$ in the low-energy spectrum, the $U(1)'$ current must be conserved. This places a constraint on the couplings in Eq.~(\ref{Lmaster}), 
\be \label{anomalyfreecondition}
c_{t_L} - c_{t_R} =-( c_{T_L} - c_{T_R}) \ .
\ee
One cannot decouple $T$ by simply setting its couplings to zero, instead they make non-decoupling contributions through the anomaly diagrams,
whose values are dictated by Eq.~(\ref{anomalyfreecondition}), and their net effect is equivalent to the Wess-Zumino terms discussed in Section~\ref{sec:lowenergyEFT}.  Note, both Eqs.~(\ref{couplingseff}) and (\ref{couplingsg}) satisfy the condition of anomaly cancellation in Eq.~(\ref{anomalyfreecondition}), which has important implications on the effects of the heavy $T$ fermion in the loop production channels of the $Z'$.

First consider the processes $q\bar q \to gZ'$ and $q g \to qZ'$. The leading Feynman diagrams for these processes are shown by those in the second row of Fig.~\ref{fig:Feynman}, both of which involve the 3-point $Z'gg$ coupling generated at loop level. Because the gluon couplings conserves $C$ parity, a generalized Furry theorem guarantees that the contribution from the vector-current coupling of $Z'$  to the loop fermion vanishes. Only the axial-current-coupling part contributes which is directly connected to the $U(1)'\otimes SU(3)_C^2$ gauge anomaly. Hereafter, we define the general effective $Z'gg$ vertex induced by a fermion $f$ loop to be, 
\be\label{eq:Effvertex}
g_s^2 {\rm Tr}(T^a T^b) \frac{c_{f_L} - c_{f_R}}{2} \Gamma_{\sigma\rho\mu}(k_1, k_2, k_3, m_f) \varepsilon^{\sigma a*}_{g}(k_1) \varepsilon^{\rho b*}_{g}(k_2) \varepsilon^{\mu*}_{Z'}(k_3) \ ,
\ee
where $a, b$ are the gluon color factors, $k_1, k_2, k_3$ are the external momentum flow (out of the loop), and the polarization vector $\varepsilon$'s indicate the combination of momenta and Lorentz indices for the gauge bosons. $\Gamma_{\sigma\rho\mu}(k_1, k_2, k_3, m_f)$ is the form factor for the fermion loop.
In the Appendix~\ref{pathintegral}, we discuss the derivation of this form factor using the path integral in the large $m_f$ limit.

Applying this to our model, and using the anomaly cancellation relation Eq.~(\ref{anomalyfreecondition}), the $Z'gg$ vertex takes the form
\be\label{formfactortT}
\frac{1}{4} g_s^2 \delta^{ab} (c_{t_L} - c_{t_R}) \left[ \Gamma_{\sigma\rho\mu}(k_1, k_2, k_3, m_t) - \Gamma_{\sigma\rho\mu}(k_1, k_2, k_3, M_T) \rule{0mm}{4mm}\right] \varepsilon^{\sigma a*}_{g}(k_1) \varepsilon^{\rho b*}_{g}(k_2) \varepsilon^{\mu*}_{Z'}(k_3) \ .
\ee
The most important feature of it is that only the difference of the two form factors (from $t$ and $T$ loops) enter in the result. It implies that one may redefine $\Gamma$ by a constant universal to all the fermions, without affecting the result.

The freedom of redefining the above form factor is closely related to the choice of consistent versus covariant anomalies discussed previously, see also~\cite{Dror:2017ehi,Dror:2017nsg, Ismail:2017ulg}.  The two are related to each other through the Wess-Zumino term. 
In the consistent anomaly case, the divergence of the form factor takes symmetric forms with respect to all the three external momenta $k_1, k_2, k_3$. 
On the other hand, for the covariant anomaly, one imposes the following gauge invariant condition for the $SU(3)_C$
\be
\label{eq:covformfactor}
\Gamma_{\sigma\rho\mu}(k_1, k_2, k_3, m_f) k_1^\sigma = \Gamma_{\sigma\rho\mu}(k_1, k_2, k_3, m_f) k_2^\rho = 0 \ .
\ee
In this case, the form factor $\Gamma_{\sigma\rho\mu}(k_1, k_2, k_3, m_f)$ takes the following explicit form~\cite{Rosenberg:1962pp}
\be\label{Rosenberg}
\frac{1}{\pi^2}\int_0^1 dx \int_0^{1-x}dy  \frac{x(x+y-1) \varepsilon_{\alpha\mu\rho\sigma} k_2^2 k_{1\alpha} - 
y(x+y-1) \varepsilon_{\alpha\mu\rho\sigma}  k_1^2 k_{2\alpha} + xy k_{3\mu} k_1^\alpha k_2^\beta \varepsilon_{\alpha\beta\rho\sigma}}{y(1-y)k_1^2 + x(1-x)k_2^2 + 2xyk_1\cdot k_2 - m_f^2} \ ,
\ee
which vanishes as  $1/m_f^2$ in the large fermion mass limit $m_f\to \infty$. In other words, the heavy $T$ fermions decouple completely and there is no need to introduce Wess-Zumino term. On the other hand, in consistent anomaly approach one chooses a momentum routing in a way that is symmetric with respect to  all three external momenta. Then the form factor $\Gamma_{\sigma\rho\mu}(k_1, k_2, k_3, m_f)$ does not vanish as $m_f\to \infty$ but instead approaches a constant. This is the explicit realization of a Wess-Zumino term from integrating out a heavy spectator fermion whose sole purpose is to cancel the $U(1)'$ anomaly. But since the vertex function in Eq.~(\ref{formfactortT}) is only proportional to the difference in the two form factors from the $t$ and the $T$ loops, the result is independent of whether one uses the covariant anomaly or the consistent anomaly approach, as it should be.  In the Appendix~\ref{pathintegral}, we derive this form factor in the large $m_f$ limit using the path integral approach.

When using a software package to perform the one-loop calculation, it is then important to keep the spectator $T$ fermions in the loop and implement the anomaly cancellation condition in Eq.~(\ref{anomalyfreecondition}), unless one can be sure that the software employs the momentum routing scheme of the covariant anomaly approach, where the $T$ fermion decouples completely. Otherwise, erroneous results will follow.

To demonstrate the importance of the preceding discussion in practice, we calculate the cross sections for $q\bar q \to gZ'$ and $q g \to qZ'$ in two ways, analytically using the above form factor, and numerically using {\tt FeynArts/FormCalc}. Not surprisingly, the calculation of {\tt FeynArts/FormCalc} does {\em not}  correspond to the covariant anomaly approach, {\it i.e.}, each $\Gamma_{\sigma\rho\mu}(k_1, k_2, k_3, m_f)$ term in {\tt FeynArts/FormCalc}  differs from Eq.~(\ref{Rosenberg}) by a universal constant in the limit $m_f\to \infty$. As a result, if the heavy fermion $T$ is neglected in the implementation of the model file, the resulting amplitudes from {\tt FeynArts/FormCalc} would not respect $SU(3)_C$ gauge invariance. Neither does it agree with the analytic calculation using the covariant anomaly approach in Eq.~(\ref{Rosenberg}). But if the heavy fermion $T$ is explicitly implemented  in {\tt FeynArts/FormCalc} with the anomaly cancellation condition in Eq.~(\ref{anomalyfreecondition}), then the three-point vertex $Z'gg$ from the two computations agree.

\begin{figure}[t]
   \centering
   \includegraphics[width=0.82\textwidth]{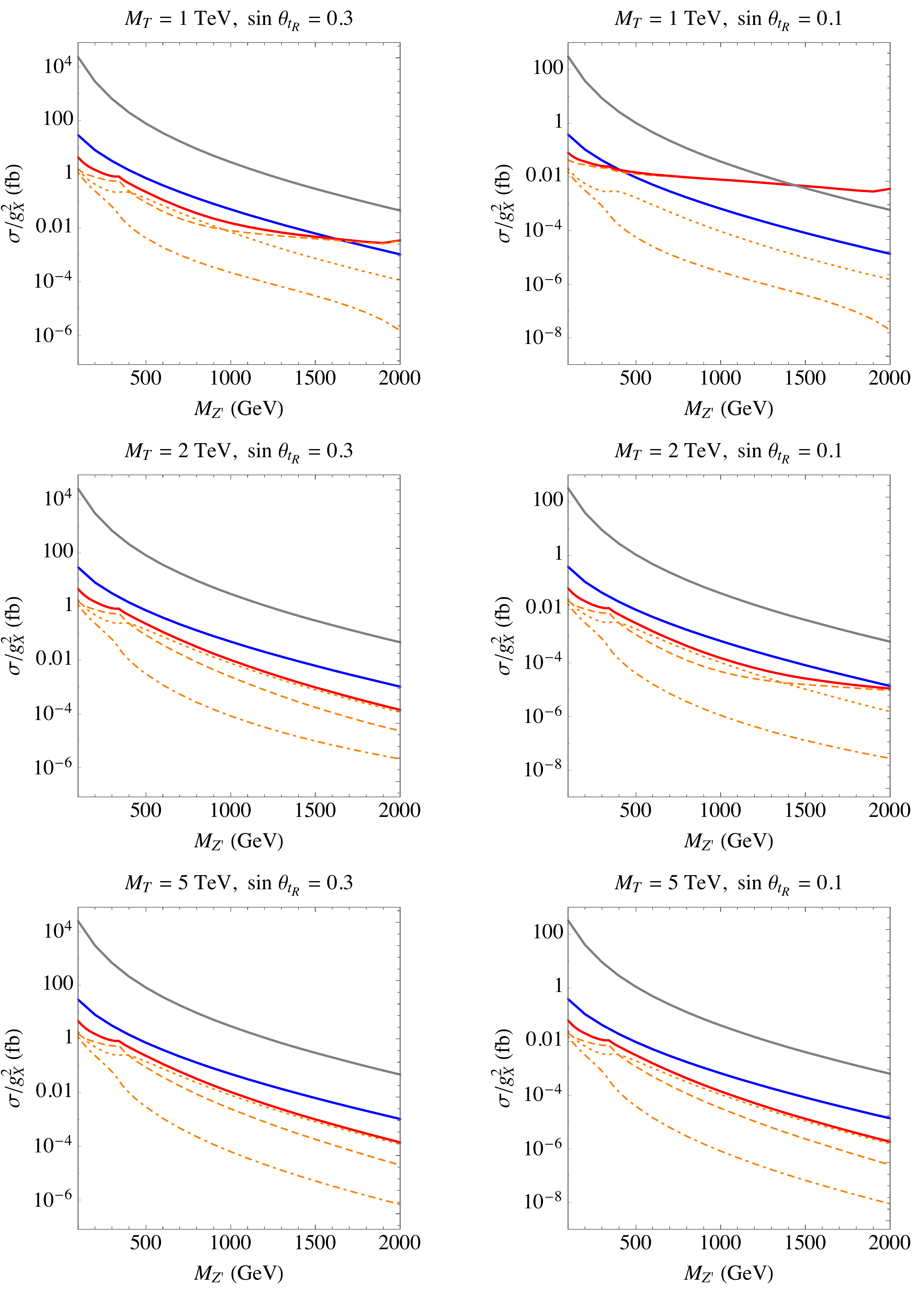} 
   \caption{$Z'$ production cross sections at 13 TeV LHC for various channels and various sets of parameters in the model we consider. {\sf Blue}: tree-level $pp\to t\bar t Z'$; {\sf Red}: sum of all loop level channels; {\sf Orange dashed}: loop-level $gg\to g Z'$; {\sf Orange dot-dashed}: loop-level $q\bar q \to gZ'$; {\sf Orange dotted}: loop-level $q g \to qZ'$ and $\bar q g \to \bar q Z'$. {\sf Grey}: sum of all loop level channels in the incomplete theory without $T$.}
   \label{fig:comparecrosssections}
\end{figure}

In Fig.~\ref{fig:comparecrosssections}, we compare the tree and loop cross sections for various couplings and masses. We have implemented the heavy $T$ fermion with the anomaly cancellation condition in the model files.  In addition, the grey curve shows the sum of loop cross sections in an incomplete model where the $SU(3)_C$ gauge invariance is lost in the $Z'gg$ vertex (\ref{eq:Effvertex}), by neglecting the $T$ fermion in the loop. Clearly, the incomplete model substantially overestimates the cross section.

\begin{figure}[t]
   \centering
   \includegraphics[width=0.9\textwidth]{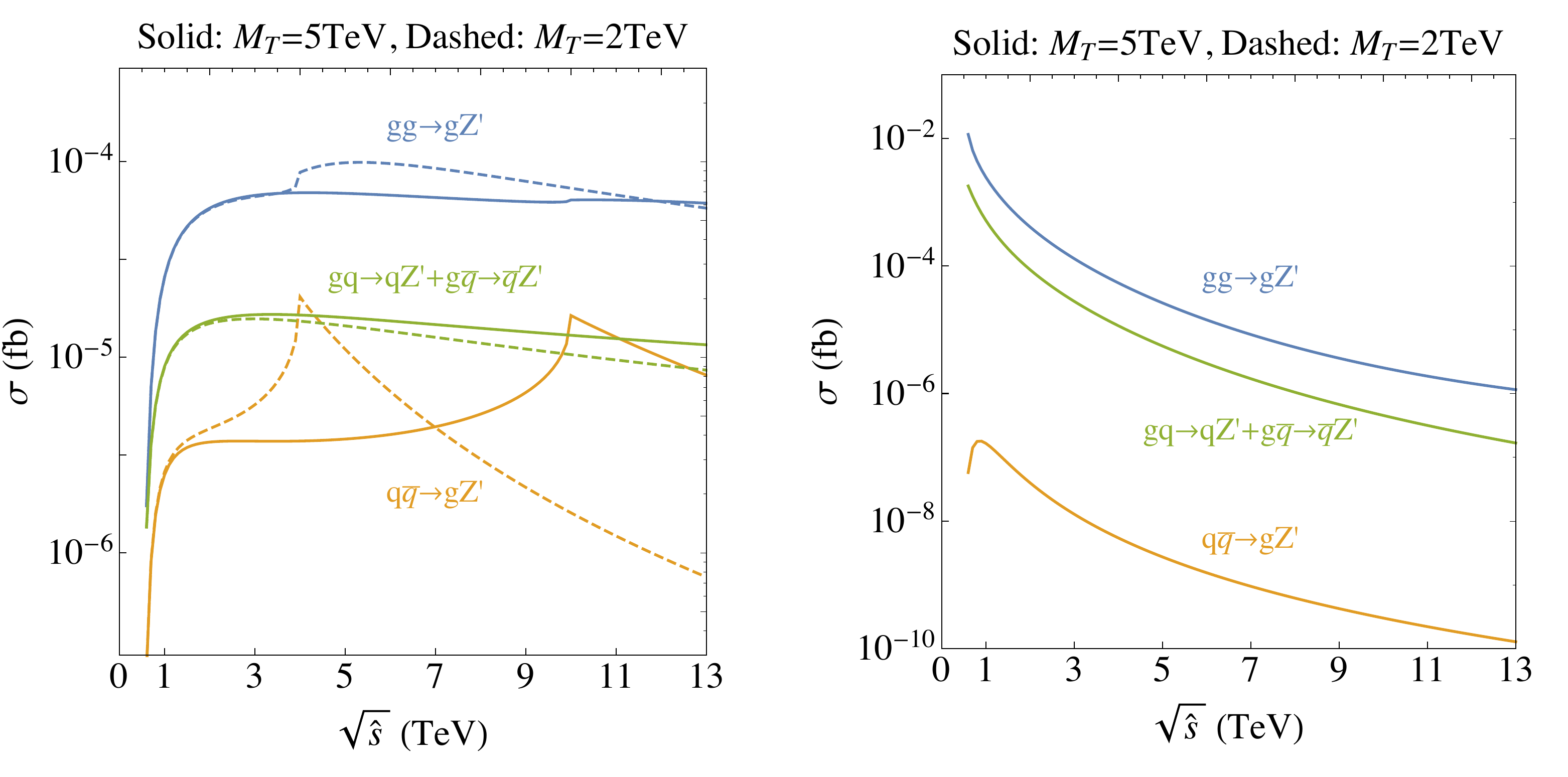}
   \caption{{\sf Left}: UV complete model with $M_{Z'}=500$\,GeV, $c_{t_L}=0$, $c_{t_R}=0.1$, $c_{T_L}=1$, $c_{T_R}=0.9$ and for two values of $M_T=5\,$TeV (solid), 2\,TeV (dashed).
   {\sf Right}: Incomplete model with $M_{Z'}=500$\,GeV, $c_{t_L}=0$, $c_{t_R}=0.1$, $c_{T_L}=0$, $c_{T_R}=0$. }
   \label{compare}
\end{figure}

To understand this large overestimate of the incomplete model consider the parton level production cross sections in Fig.~\ref{compare}.  We show the parton level $Z'$ production cross sections via loops, for various initial states, as a function of the center-of-mass energy $\sqrt{s}$, in the UV complete (LH plot) versus incomplete model (RH plot).
With the chosen parameters, the $gg\to gZ'$ channel is dominating the production. The parton level cross section predicted by the incomplete model is highly peaked in the low $\sqrt{s}$ regime. In contrast, there is a non-trivial cancellation between the $t$ and $T$ loops contributions seen in the UV complete model.
This is the reason why after the convoluting integral with the PDF's, the incomplete model yields a much larger cross section than the UV complete model (see Fig.~\ref{fig:comparecrosssections}). This (lack of) cancellation explains the difference between our results and those appearing earlier \cite{Greiner:2014qna,Cox:2015afa,Kim:2016plm}.

For the channels whose contributions are mainly from anomaly diagrams, we also find the asymptotic behavior of their parton level cross sections,
\be
\sigma \propto \left\{  
\begin{array}{ll}
{1}/{s}, &\hspace{0.6cm} \sqrt{s} \gg M_{T}, M_{Z'}, m_t \\
{1}/{M_{Z'}^2}, &\hspace{0.6cm} M_T \gg \sqrt{s} \gg M_{Z'}, m_t
\end{array}
\right.
\ee
In the second case with a superheavy $T$, the scaling behavior is because it is predominantly the longitudinal component of the $Z'$ that is produced~\cite{Dror:2017ehi,Dror:2017nsg}. This explains why the cross section plateaus as function of $\sqrt{s}$, up to the running of $\alpha_s$.

Finally, we comment on the $gg\to gZ'$ channel. The calculation of this process is more complicated because it involves both triangle and box diagrams, see Figure~\ref{fig:Feynman}. 
Due to the non-abelian nature of $SU(3)_C$, the box diagram also has a contribution from the anomaly.
The above discussions will have the same impact on the calculations of these contributions.
In addition, the vector-current $Z'$-fermion interaction could also make a contribution to the box diagram, which in the heavy fermion limit, corresponds to
an Euler-Heisenberg-like effective operator.
In practice, we calculate the $gg\to gZ'$ cross section numerically using {\tt FeynArts/FormCalc}.

In the case of $\sqrt{\hat s} > 2M_T$, we verified numerically that the case where $Z'$ couples to $T$ through a vector-current gives the dominant contribution to $gg\to gZ'$ over the anomaly-related ones (with a small $\theta_R$). This explains why in the $M_T=1\,$TeV case (the top two plots in Fig.~\ref{fig:comparecrosssections}), when $M_{Z'} \to 2\,$TeV (remember that $\sqrt{\hat s}\geq M_{Z'}$) the two orange dashed curves take approximately the same value.
In this case, the box diagram involving $T$ is the most important (there is no mass suppression at high enough energy) and the vector-current coupling is $(c_{T_L}+c_{T_R})/2$ is always $\sim1$ for small enough $\theta_R$.
In the other plots of Fig.~\ref{fig:comparecrosssections}, we have $M_{Z'} \ll 4 M_T$ thus the region with $\sqrt{\hat s} \ll 2M_T$ gives the most important contribution to the LHC cross section, and the anomaly-related diagrams are the most important. 

\subsection{Our recipe for loop process calculations}

To summarize, we provide the following options for properly doing the calculation of loop-level $Z'$ production in a way that respects the $SU(3)_C$ gauge invariance:
\vspace{0.1cm}
\begin{itemize}
\item Do the calculation with both $t$ and $T$ running in the loop. Moreover, we must ensure their couplings satisfy the anomaly free condition in Eq.~(\ref{anomalyfreecondition}).\vspace{0.1cm}
\item Calculate the diagrams with only top quark running inside  the loop using the form factor of the covariant anomaly case, Eq.~(\ref{eq:covformfactor}). Then the heavy $T$ effects can be safely decoupled.\vspace{0.1cm}
\item Calculate the top quark loop and properly include the appropriate Wess-Zumino term such that the $SU(3)_C$ gauge invariance is maintained in the amplitude.
\end{itemize}

\section{LHC bound on the top-philic $Z'$ using multi-top final states}\label{sec:LHCtopbounds}

In this section, we classify the possible final states related to top-philic $Z'$ production at the LHC and estimate the current bounds.  Our discussion is based on Eq.~(\ref{Lmaster}) in the effective $Z'$ model, where the $Z'$ only has couplings to the top quark and the heavy top partner, but the results also apply to the gauged top model.
We assume the mass of $Z'$ lies in the range $2m_t < M_{Z'} < M_T - m_t$, so that the decays $Z'\rightarrow t\bar{t}$ and $T\rightarrow Z' t$ produce onshell decay products.  Furthermore, for simplicity we assume that the decay kinematics are such that the top quarks are well separated. 

\subsection*{$t\bar t$ {\bf final states}}
As discussed in the previous section, a top-philic $Z'$ boson can be produced at loop level in association with a jet. The $Z'$ then decays into $t\bar t$ and it can be searched for as a $t\bar t$ resonance. A recent study by the ATLAS collaboration based on a fraction (3.2\,fb$^{-1}$) of existing the 13\,TeV data~\cite{TheATLAScollaboration:2016wfb} sets an upper limit on the $Z'$ production cross section ranging from 300\,pb to 0.2\,pb for $Z'$ mass between 500\,GeV and 2\,TeV. This is a rather weak bound compared to the $Z'+j$ production cross sections derived in Fig.~\ref{fig:comparecrosssections}. 

\subsection*{{\bf Four top final states}}
The $Z'$ can be produced at tree level in association with $t\bar t$.  After its decay this leads to the four top quark final state $t\bar{t}t\bar{t}$. A very recent study of the four top channel by the CMS collaboration, using 35.9\,fb$^{-1}$ of data~\cite{Sirunyan:2017roi}, measures $\sigma_{4t}=16.9^{+13.8}_{-11.4}\,$fb, while the SM prediction is $\sigma_{4t}^{\rm SM}=12.2\,$fb.
This leads to an upper bound on the new physics 4-top production cross section to be $\sigma_{4t}^{\rm NP} \le 32.3\,$fb.
This is a relatively weak bound when compared to the results of Fig.~\ref{fig:comparecrosssections}.  
For instance, at threshold for $Z'$ to decay to $t\bar{t}$ the coupling is bounded by $c_{t_R}=g_X \sin^2\theta_R \ltap 1$.  In the future, with higher LHC luminosities, it will place a non-trivial limit on the model parameter space. We will further quantify the future prospects in the next section and in particular in Fig.~\ref{fig:combined}.

\subsection*{{\bf Six top final states}}
If the heavy top partner is pair produced at the LHC, each will decay into $Z'$ and $t$ (or $\bar t$) as dictated by Eq.~(\ref{Lmaster}). 
Finally, after $Z'$ decays into $t\bar t$, the final state will contain six top quarks. Currently, there is no dedicated search for a six top quark final state at the LHC.
However, it is possible for the decay products of the six top quarks to fall into the signal regions of the above four top search.
In~\cite{Sirunyan:2017roi}, the number of events have been measured in 8 signal regions characterized by two or three charged leptons ($e$ or $\mu$) and two to four $b$ jets.
In each signal region the number of observed events, the number predicted by SM four top, and the number predicted from non-four top processes are reported. 
Based on the the SM four top cross section, the branching fraction of four tops into each signal region, and a b-tagging efficiency of $60\%$ we estimate the signal efficiency factor for a true four top final state to appear in each signal region.  For simplicity, we assume that these efficiency factors are independent of the $p_T$ of the leptons and $b$ jets and that they can be applied to the decay products of $T$.  Thus, for a given six top quark production cross section we can estimate the number of events expected in each of the signal regions of the four top quark analysis.
Comparing with the existing data, we derive an upper limit on the six top quark production cross section $\sigma_{6t} \lesssim 3\,$fb.  Since the vectorlike top partners are mainly produced via QCD interactions, this in turn requires $M_T>1.3$\,TeV~\cite{Czakon:2011xx,Fox:2011qd,Alwall:2011uj}.  This bound can be weakened if there are additional exotic decay modes of the $T$ \cite{Dobrescu:2016pda}.

In addition to the searches described above, which are tailored to search for top quarks in the final state, these multi-top final states will also appear in  searches for supersymmetry in multilepton final states.   If the $Z'$ also decays to leptons the multilepton rates will increase further.  A full analysis of the many signal regions in these searches \cite{Sirunyan:2017hvp,Sirunyan:2017lae} is beyond the scope of this work, but may provide interesting constraints.

\section{Application to recent LHCb excesses}\label{sec:pheno}

In this section, we apply the general discussions in the previous sections to a model with $Z'$ boson and vectorlike quarks, which have been introduced for understanding the
recent anomalies in $b\to s\mu^+\mu^-$ transition observables at LHCb. For recent global analysis of the effects in LHCb measurements of $R_K$, $R_{K^*}$, $P_5'$ as well as other flavor observables at LHCb and BaBar, see {\it e.g.},~\cite{Descotes-Genon:2015uva, Ciuchini:2017mik, Capdevila:2017bsm,DAmico:2017mtc,Altmannshofer:2017yso,Geng:2017svp, Hiller:2017bzc, Celis:2017doq, Ghosh:2017ber, Alok:2017sui}.  As pointed out in~\cite{Kamenik:2017tnu}, this anomaly can be explained without new sources of flavor violation with the addition of a new gauge boson that couples at tree level to only right-handed top quarks and second generations leptons.  At low energies the relevant couplings of the $Z'$ gauge boson are,
\be
\label{eq:IRLagrangian}
c_{t_R} \bar{t} \slashed{Z'} P_R t + \bar{\mu} \slashed{Z'}  (c_L P_L + c_E P_R) \mu +  c_L \bar{\nu_\mu} \slashed{Z'}  P_L \nu_\mu~.
\ee
These interactions alone make the $U(1)'$ anomalous and the theory must be UV completed.  We consider two possible simple completions, corresponding to the two models discussed in Section~\ref{sec:twomodels}.  The effective $Z'$ model, in which we introduce vectorlike top partners charged under the $U(1)'$ and turn on their mixings with the SM top quark with the VEV of a new scalar field.  And the gauged top model, where the SM right-handed top quark is directly charged under the $U(1)'$ and the new fermions that cancel the anomaly are chiral under the $U(1)'$. These additional states may be sufficiently heavy that they only show up in loop processes.  Similarly, the effective couplings of the $Z'$ to $\mu$ and $\nu_\mu$ can be generated by introducing heavy vector-like leptonic partners.  Since they are not colored the constraints on their masses, and mixings with SM partners, are weaker.

\subsection{B-physics anomalies}

Global fits~\cite{Capdevila:2017bsm,DAmico:2017mtc,Altmannshofer:2017yso,Geng:2017svp,Ciuchini:2017mik} to the LHCb excesses, as well as various sets of other B-physics observables, show that they can be fit with two four-fermion operators,
\be\label{eq:C9C10}
\mathcal{H}=-\frac{4G_F}{\sqrt2} V_{tb}V_{ts}^* \frac{\alpha_{em}}{4\pi} 
\Big(
C_9^\mu \mathcal{O}_9^\mu 
+ C_{10}^\mu \mathcal{O}_{10}^\mu
\Big) + {\rm h.c.}~,
\ee
with $\mathcal{O}^\ell_9=\left(\bar{s}\gamma^\mu P_L b\right)\left(\ell \gamma_\mu \ell\right)$ and $\mathcal{O}^\ell_{10}=\left(\bar{s}\gamma^\mu P_L b\right)\left(\ell \gamma_\mu \gamma_5 \ell\right)$, and that the best fit region lives in the quadrant with $\delta C_9<0$ and $\delta C_{10}>0$.    
An interesting feature of the $Z'$ models is that the couplings of the $Z'$ do not change the quark flavors.  Any FCNC effect induced by the $Z'$ must occur at loop level, through diagrams involving additional sources of flavor violation \eg\ $W^\pm$ or $H^\pm$ bosons running in loops~\cite{Kamenik:2017tnu}. This not only reduces the number of new parameters but also requires the $Z'$ mass to be not far above the electroweak scale in order to give a significant contribution to the effective operators, $\mathcal{O}_9$
and 
$\mathcal{O}_{10}$. 
It is also worthwhile pointing out that because $\bar s \gamma^\mu P_L b$ is a conserved current if the quark masses are omitted, the $\mathcal{O}_{9,10}$ operators have no anomalous dimensions from QCD radiative corrections between the scale where these effective operators are generated and the $B$-meson mass scale.

\subsubsection*{Effective $Z'$}

\begin{wrapfigure}[13]{l}{0.3\textwidth}
\centering\includegraphics[width=0.26\textwidth]{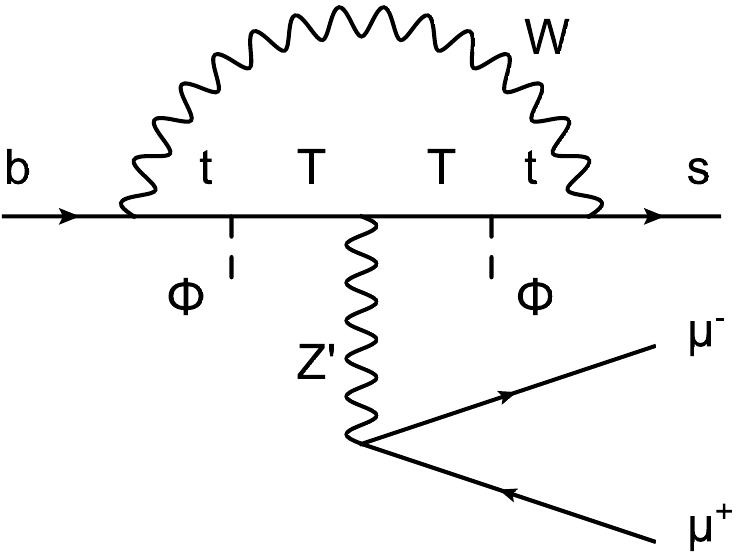} 
\caption{Feynman diagram for the effective $Z'$ model contribution to $b\to s \mu\mu$.}
   \label{fig:Bloop}
\end{wrapfigure}

If we embed the low-energy Lagrangian Eq.~(\ref{eq:IRLagrangian}) in the context of the effective $Z'$ model described in Subsection~\ref{effZ'}, the couplings of the $Z'$ to the SM top quark and the heavy vectorlike fermion $T$ take the approximate form,
\bea\label{eq:c's}
\begin{split}
&c_{t_R} = g_X\sin^2\theta_R\ , \ \ \ c_{t_L} \simeq0 \ ,  \\
&c_{T_R} = g_X\cos^2\theta_R \ , \ \ \  c_{T_L} \simeq g_X \ ,
\end{split}
\eea
where $\theta_R$ is the mixing angle between the right-handed $t_R$ and $T_R$ and we have taken $M_T\gg m_t$. 
The loop contribution is shown in Fig.~\ref{fig:Bloop}. It is finite because the $Z'bs$ coupling originates from a dimension 6 operator in the complete theory, $(\bar s\gamma^\mu P_L b)(\Phi^*\overleftrightarrow{D}_\mu\Phi)$.
The corrections to the Wilson coefficients are,
\be\label{eq:C9&10}
\delta C_{9,10} = \frac{g_X\sin^2\theta_R\left(c_E\pm c_L\right) m_t^2}{4 M_{Z'}^2 e^2} 
\left[ \ln \frac{M_T^2}{m_t^2} + \frac{3 M_W^2}{m_t^2 - M_W^2} - \frac{3M_W^4}{(m_t^2 - M_W^2)^2}\ln\frac{m_t^2}{M_W^2}
 \right] +\mathcal{O}\left( \frac{m_t^2}{M_T^2} \right) \ .
\ee

The best fit region favors $c_L\gg c_E$ in the low-energy Lagrangian Eq.~(\ref{eq:IRLagrangian}).
Assuming $c_E=0$, the region of parameter space that could account for the $B$-physics anomalies is shown by the green regions in Fig.~\ref{fig:zupanplots}.
Fig.~\ref{fig:zupanplots} shows the $B$-physics favored parameter space in the $g_X$-$M_{Z'}$ plane, with two sets of values of $\theta_R$ and $c_L$. For the coupling $g_X$ to remain perturbative, we must resort to sub-TeV $Z'$ mass and sizable mixing angle $\theta_R\gtrsim 0.1$.
Using the parameter relation Eq.~(\ref{eq:pararelations}) we find that the vectorlike fermion $T$ cannot be arbitrarily heavy, and in turn, the logarithmic factor in Eq.~(\ref{eq:C9&10}) is not large, and the finite correction terms in the square bracket are also important.

\subsubsection*{Gauged top model}
On the other hand, if we embed the low-energy Lagrangian Eq.~(\ref{eq:IRLagrangian}) in the context of the gauged top model described in Subsection~\ref{sec:gaugedtop}, the $Z'$ coupling to the top quark and the muon is tied closely to each other. In the $\mu\to0$ limit, the heavy vectorlike top do not mix with the light one and do not contribute to the $b\to s$ transition. The relevant couplings are
\bea\label{eq:c'sothermodel}
\begin{split}
c_{t_R} = c_L = g_X\ , \ \ \ c_{t_L} = 0 \ ,  \ \ \  c_{T_L} = c_E=0 \ .
\end{split}
\eea
The calculation of the new contribution to $\delta C_{9,10}$ in this model is more complicated, and since $c_L\gg c_E$ we have fixed $c_E=0$ for simplicity. 
The result is of course finite, but there is a non-trivial cancellation among the UV divergent parts.
We give more details of this calculation in the Appendix~\ref{app:gaugedtop}.
In the heavy $H^{\pm}$ limit, the Wilson coefficients of interest to the $b\to s\mu\mu$ process are,
\be\label{eq:C9&10gaugedtop}
\delta C_{9,10} = \mp\frac{g_X^2 m_t^2 \cos^2\beta (1-\sin^2\beta \cos2\theta_W)}{4 M_{Z'}^2 e^2} 
\left[ \ln \frac{M_{H^\pm}^2}{m_t^2} - \frac{m_t^2 - 4 M_W^2}{m_t^2 - M_W^2} - \frac{3M_W^4}{(m_t^2 - M_W^2)^2}\ln\frac{m_t^2}{M_W^2}
 \right] +\mathcal{O}\left( \frac{m_t^2}{M_{H^\pm}^2} \right) \ ,
\ee
In the gauged top model, with $c_E=0$, we always have the relation that $\delta C_9=-\delta C_{10}$. We set $M_{H^\pm}=10\,$TeV in our calculation, which is large enough to suppress all the $\mathcal{O}\left( {m_t^2}/{M_H^2} \right)$ correction terms. 
Because the gauged top model is a two-Higgs doublet model, we choose to work in the alignment limit which is most consistent with the LHC Higgs rate measurements~\cite{Chen:2013rba,Craig:2015jba,Chen:2015gaa}.
The $B$-physics favored parameter space in the $g_X$-$M_{Z'}$ plane is shown in Fig.~\ref{fig:zupanplots2}.

It is also worth pointing out that the above calculation is done by assuming no kinetic mixing between the $U(1)'$ and hypercharge gauge bosons, which is a marginal and gauge invariant operator in the complete theory. Dressing the SM penguin diagrams with this mixing and the $Z'$-muon coupling will make additional contribution to $C_{9, 10}$. If the effective $Z'$ is further embedded in more unified models, it is conceivable that the kinetic mixing vanishes at high scale and is only generated through the running effect. In this case, its contribution to $C_{9, 10}$ would be at two-loop level and negligible compared to the one-loop contributions given in Eq.~(\ref{eq:C9&10}).

\subsection{LHC dimuon resonance search for $Z'$}

\begin{wrapfigure}{l}{0.3\textwidth}
\centering\includegraphics[width=0.26\textwidth]{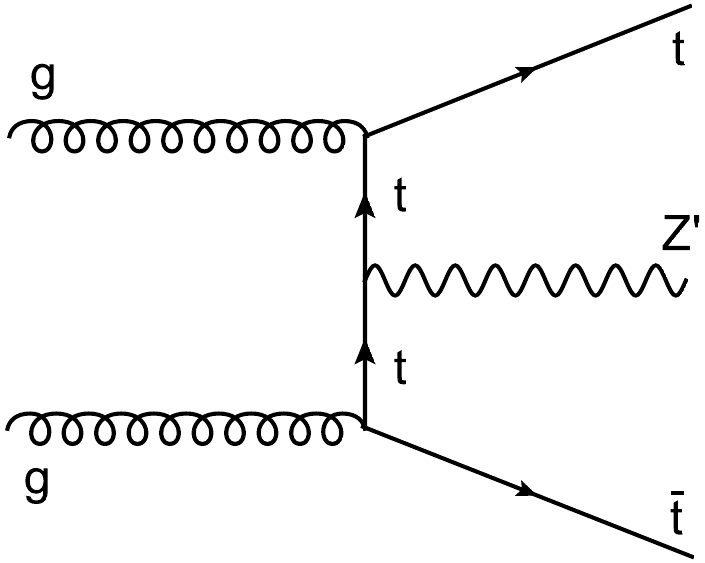} 
\caption{Feynman diagram for the effective $Z'$ model contribution to $b\to s \mu\mu$.}
   \label{fig:ggtTZp}
\end{wrapfigure}

\begin{figure}[t]
   \centering
    \includegraphics[width=0.45\textwidth]{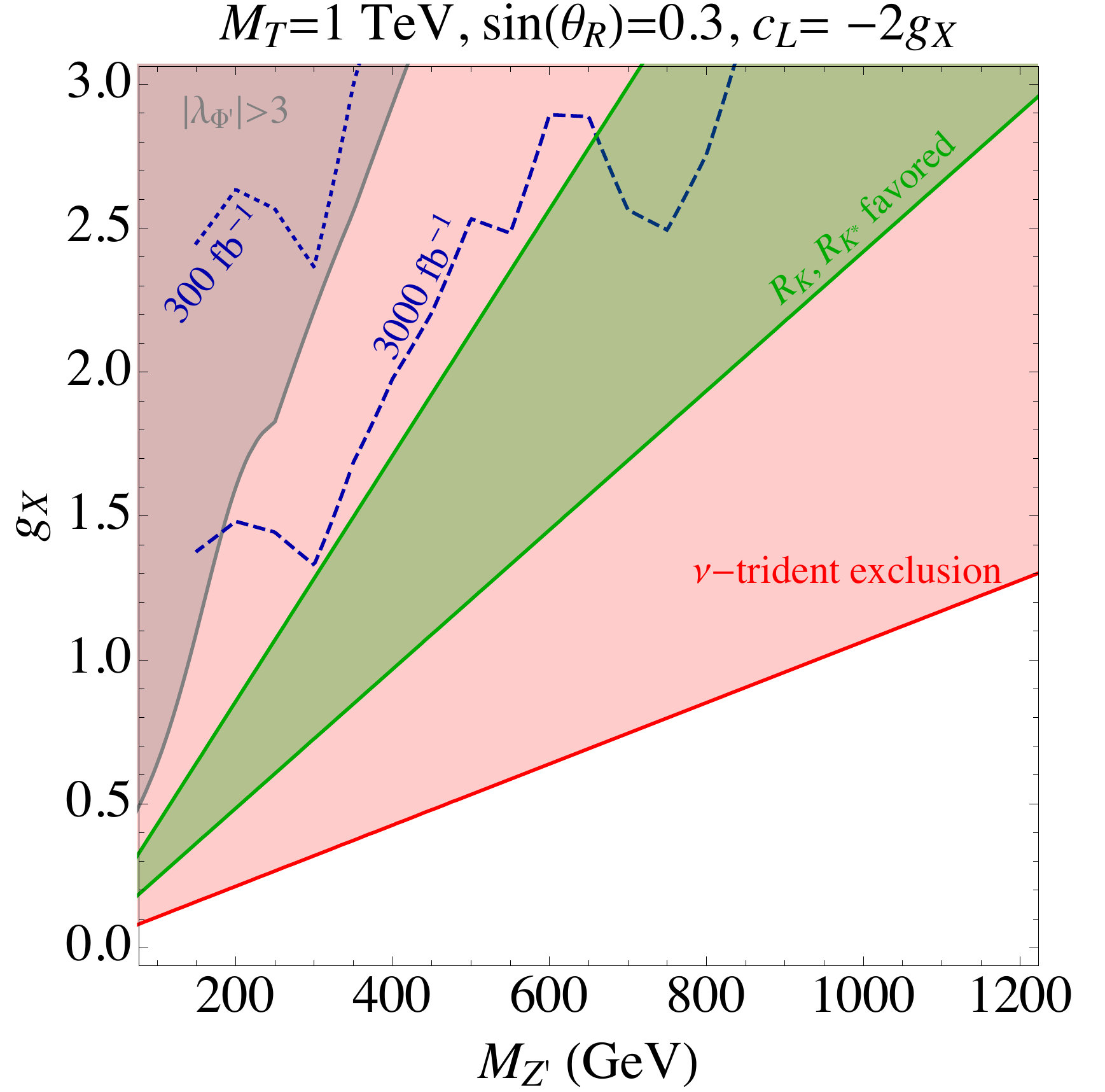}
    \includegraphics[width=0.45\textwidth]{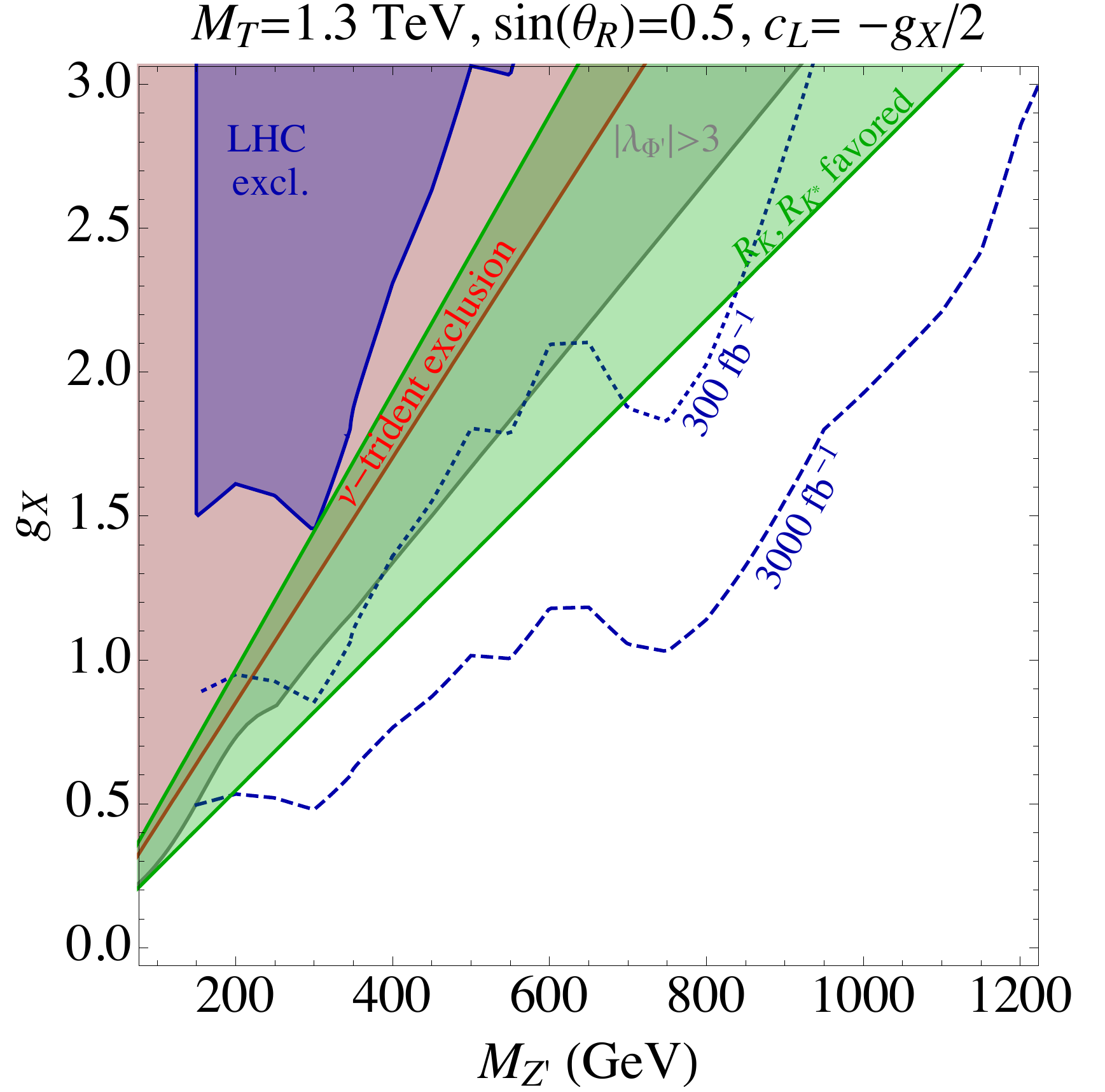}
   \caption{Favored regions and constraints on the parameter space (the $g_X$-$M_{Z'}$ plane) of the effective $Z'$ model, defined in Eq.~(\ref{eq:IRLagrangian}) and (\ref{eq:c's}), with $M_T=1\,$TeV, $\sin\theta_R=0.3$, $c_L=2 g_X$, $c_E=0$ (left); $M_T=1.3\,$TeV, $\sin\theta_R=0.5$, $c_L= g_X/2$, $c_E=0$ (right). The green region in each plot is favored for explaining the LHCb anomalies at $2\sigma$ C.L.~\cite{Altmannshofer:2017yso}, which corresponds to $C^\mu_9 = - C^\mu_{10}$ and $-1 < C^\mu_9 < -0.32$. The red shaded regions are excluded by the neutrino trident production measurement. The blue shaded region is excluded by the LHC dimuon resonance search. The dashed (dotted) blue curve corresponds to the future LHC reach with an integrated luminosity 300 (3000)/fb.}
   \label{fig:zupanplots}
\end{figure}

An important message from Fig.~\ref{fig:zupanplots} is that the mass of $Z'$ in this model must lie below $\sim {\rm TeV}$ scale in order to account for the $B$-physics anomalies. Because the $Z'$ must couple to muons, the dilepton resonance search at the LHC~\cite{Aaboud:2017buh} serves as one of the leading measurements to test such an explanation.

As discussed in Section~\ref{sec:Z'prod}, there are two important production channels of a top-philic $Z'$ boson at the LHC.
One occurs at tree-level, where the $Z'$ is produced in association with $t\bar t$. A representative Feynman diagram for this process is shown in Fig.~\ref{fig:ggtTZp}.
The second channel is to produce the $Z'$ at loop level in association with a jet, as shown by Fig.~\ref{fig:Feynman}.
As we have discussed in great detail, the anomaly cancellation plays an important role in such loop processes and the heavy $T$ must be taken into account
when calculating the cross sections using {\tt FeynArts} and {\tt FormCalc}.
We take both production channels into account in our analysis.
The comparisons in Fig.~\ref{fig:comparecrosssections} shows that for most of the parameter space with $M_T>M_{Z'}$, the tree-level cross section 
is higher than the loop-level one by a factor of a few.\footnote{There are also loop production channels of the $Z'$ which involve electroweak interactions, such as $t j Z'$, $t W Z'$ and $WWZ'$. We find their cross sections are all negligibly small.}

On the other hand, although the dimuon resonance is an inclusive search which allows additional activity in each event, when the $Z'$ is produced together with $t\bar t$ the top quark decay products could reduce the selection efficiency of isolated muons in the final states. In order to estimate the efficiency, we use {\tt MadGraph}~\cite{Alwall:2014hca} to simulate the $Z't\bar t$ events, run the hadronization with {\tt PYTHIA} and the detector simulation with {\tt Delphes} using the default isolation criterion.
Requiring that the two leading isolated, opposite-sign muons to be reconstructed, we find the selection efficiency for this channel is roughly 0.4.
In contrast, the efficiency for the loop produced $Z'$ channels is almost 1.
With these efficiency factors taken into account, we find the contribution from loop level $Z'+j$ production can be as large as 50\% of that from the tree level $Z't\bar t$ channel.

The three dominant decay modes of the $Z'$ boson are,
\bea\label{eq:Z'decay}
\begin{split}
&\Gamma(Z'\rightarrow \mu^+\mu^-) = \frac{c_L^2+c_E^2}{24\pi}M_{Z'}~,\\
&\Gamma(Z'\rightarrow \bar{\nu}\nu) = \frac{c_L^2}{24\pi}M_{Z'}~,\\
&\Gamma(Z'\rightarrow \bar{t}t) = \frac{c_{t_R}^2}{8\pi}\left(1-\frac{m_t^2}{M_{Z'}^2}\right)\sqrt{1-4\frac{m_t^2}{M_{Z'}^2}}M_{Z'} \ .
\end{split}
\eea

The LHC dimuon constraint amounts to require that $\sigma (pp\to Z' +X) \times {\rm Br}(Z'\to \mu^+\mu^-) \times {\rm efficiency}$ to be below the upper bound provided in~\cite{Aaboud:2017buh}. In Fig.~\ref{fig:zupanplots}, the blue shaded regions are excluded by the present LHC result. 
The bound is stronger in the right plot because of the $Z'$ is more abundantly produced at LHC with the larger value of $\theta_R$. 
The production cross section is proportional to $\sin^4\theta_R$. 
Assuming the sensitivity scales as $\propto \mathcal{L}^{-1/2}$, we estimate the future LHC reach with integrated luminosity equal to 300 (3000)\,fb$^{-1}$ and show the expect reach using the blue dashed (dotted) curves.
Interestingly, the future high luminosity running of LHC could potentially cover much of the remaining region able to explain the present $B$ physics anomalies.

\subsection{$\nu$ trident production}

\begin{wrapfigure}[13]{l}{0.3\textwidth}
\centering\includegraphics[width=0.26\textwidth]{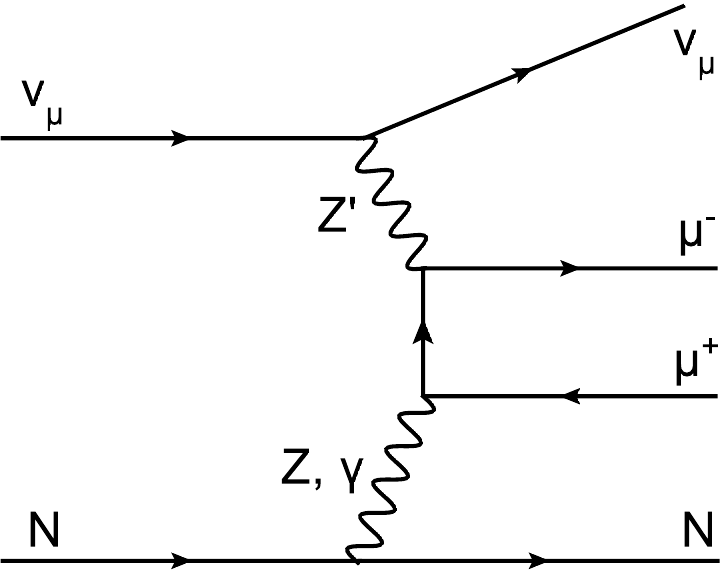} 
\caption{Feynman diagram for the effective $Z'$ model contribution to $b\to s \mu\mu$.}
   \label{fig:trident}
\end{wrapfigure}

With $c_L\neq0$, the $Z'$ necessarily couples to neutrinos and it can contribute to the neutrino trident production process $\nu N\rightarrow N \nu \mu^+\mu^-$, as shown by Fig.~\ref{fig:trident}.  The rate for this process has been measured (at CHARM-II and CCFR) to be near the SM value. The ratio of the trident cross section in the model Eq.~(\ref{eq:IRLagrangian}) to that in the SM, is given by~\cite{Altmannshofer:2014pba},
\be
\frac{\sigma_{Z'}}{\sigma_{SM}} =\frac{\left(\frac{c_L(c_L+c_E)}{M_{Z'}^2}+\sqrt{2}(1+4 s_W^2) G_F\right)^2 + \left(\frac{c_L(c_L-c_E)}{M_{Z'}^2}+\sqrt{2}G_F\right)^2}{2G_F^2\left(1+(1+4s_W^2)^2\right)} \ .
\ee
We require that this ratio lies within $2\sigma$ of the observed value \ie\ $\sigma_{Z'}/\sigma_{SM}<1.38$. In Fig.~\ref{fig:zupanplots}, the parameter space excluded by the trident observation corresponds to the red regions. Here, we find an interesting interplay between the LHC dimuon resonance search and the $\nu$ trident production measurement. In the left plot, we take a relatively larger $Z'$-muon (and neutrino) coupling $c_L$ and in this case, the trident bound completely excludes the $B$-physics favored region. In the right plot, we reduce $c_L$ in order to evade the trident bound but increase the $Z'$-top coupling (through the parameter $\theta_R$), so that the $B$-physics favored region remains. In this case, the LHC dimuon search bound gets stronger and the present data have already excluded part of the favored region.
The further running of LHC with slightly higher luminosity will enable us to either discover the $Z'$ or exclude this model as an explanation for $R_K, R_K^*$.

\subsection{Electroweak precision constraints on $Z-Z'$ mixing}\label{ZZpMix}

\begin{figure}[t]
   \centering
    \includegraphics[width=0.5\textwidth]{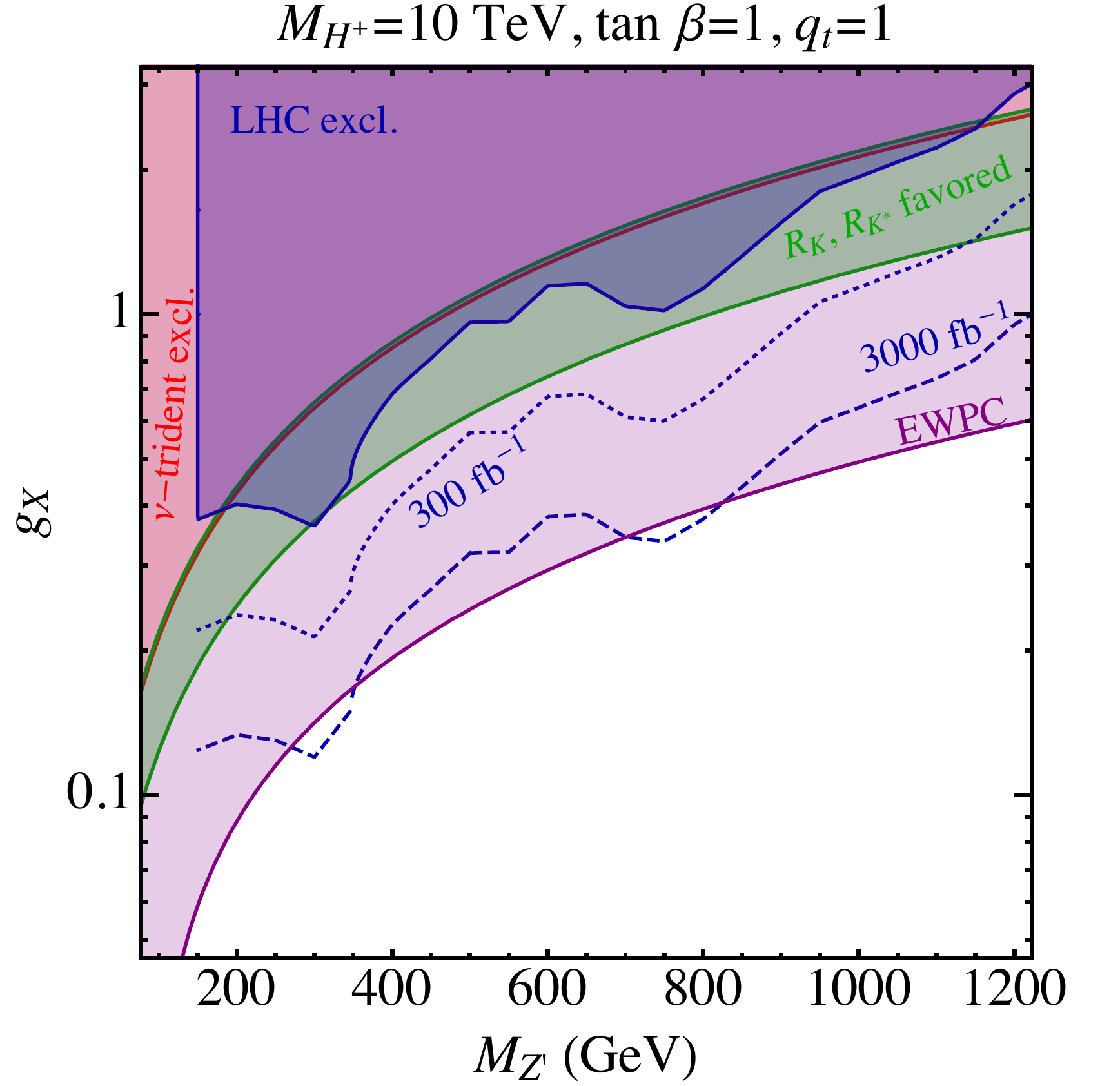}
   \caption{Favored regions and constraints on the parameter space (the $g_X$-$M_{Z'}$ plane) of the gauged top model, with $m_{H^\pm}=10\,$TeV and $\tan\beta=1$.
The green region in each plot is favored for explaining the LHCb anomalies, which corresponds to $-1 < C^\mu_9 = - C^\mu_{10} < -0.32$ at $2\sigma$ C.L.~\cite{Altmannshofer:2017yso}. The magenta region is excluded by the LEP electroweak precision meansurement.
The red shaded regions are excluded by the neutrino trident production measurement. The blue shaded region is excluded by the LHC dimuon resonance search. The dashed (dotted) blue curve corresponds to the future LHC reach with an integrated luminosity 300 (3000)/fb.}
   \label{fig:zupanplots2}
\end{figure}

In the gauged top model, the VEV of the Higgs doublet which is charged under both SM and the new $U(1)'$ yields a tree level mixing between the $Z$ and $Z'$ bosons.
Their mass matrix has been shown in Eq.~(\ref{eq:ZZ'mixing}) (see also Eq.~(\ref{eq:Mgauge})). As a consequence, the $Z$ boson mass is shifted, while the $W$ boson mass remains SM-like. This gives a contribution to the $\Delta \rho$ parameter. The current constraint on $\rho$ based on a global fit with the $S, T, U$ parameters is~\cite{PDG},
$\rho = 1.0006 \pm 0.0009$.
As shown in Fig.~\ref{fig:zupanplots2}, this sets a strong constraint on the gauged top model and all the parameter space for explaining the $B$-physics anomalies have been excluded.

\subsection{Additional constraints}

There are further constraints on the model, which could be important if we vary the parameters beyond the scope of Fig.~\ref{fig:zupanplots}. 

\vspace{0.2cm}

\begin{wrapfigure}[8]{l}{0.3\textwidth}
\centering\includegraphics[width=0.26\textwidth]{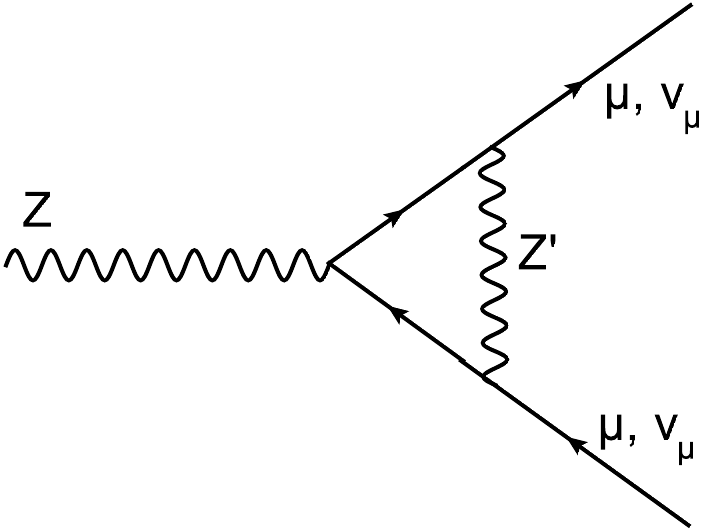} 
\caption{Feynman diagram for the effective $Z'$ model contribution to $b\to s \mu\mu$.}
   \label{fig:LEPZpole}
\end{wrapfigure}

\noindent{\it Lepton flavor universality of $Z$ couplings at LEP}. \ \  
At one loop level, the exchange of $Z'$ gives a vertex corrections to $Z$-boson coupling to the muon and its neutrino, as shown by Fig.~\ref{fig:LEPZpole}. Such a radiative correction can be constrained by the measurement at the $Z$-pole at LEP-II. 
There are 7 measured $Z$-boson couplings~\cite{ALEPH:2005ab}, among which 3 receive additional radiative corrections due to $Z'$ exchange in this model~\cite{Altmannshofer:2016brv},\\[-15pt]
\bea
g_{L_\nu} &=& \left((1 + \frac{2 c_L^2}{48 \pi^2} F\left(\frac{M_Z^2}{M_{Z'}^2}\right)\right)(g_{R_e}-g_{L_e})\nonumber \\
g_{L_\mu} &=& \left((1 + \frac{c_L^2}{16 \pi^2} F\left(\frac{M_Z^2}{M_{Z'}^2}\right)\right)g_{L_e}\\
g_{R_\mu} &=& \left((1 + \frac{c_E^2}{16 \pi^2} F\left(\frac{M_Z^2}{M_{Z'}^2}\right)\right)g_{R_e}\nonumber \ .
\eea
With $F(x) =-\frac{4+7x}{2x}+\frac{2+3x}{x}\log x-\frac{2(1+x)^2}{x^2}\left(\log x \log(1+x) + \mathrm{Li}_2 (-x)\right)$ $\to (11-6\log x)x/9+\mathcal{O}(x^2)$.  
Because of the flavor structure of the model, there are no corrections to the other $Z$-boson couplings, $g_{R_e}, g_{L_e}, g_{R_\tau}, g_{L_e\tau}$, and since the SM predictions fits the LEP-II data very well, we simplify our analysis assuming these couplings to be very close to their observed values.
To construct the $\chi^2$ for the fit, we use the correlation matrix presented in Table 7.7 of~\cite{ALEPH:2005ab}. 
When presenting constraints in Fig.~\ref{fig:combined} we hold $M_{Z'}$ fixed so there is one degree of freedom in the fit and the $95\%$ CL exclusion corresponds to $\Delta \chi^2=3.84$.  The SM limit with $c_L=c_E=0$ and $\chi^2=2.05$ gives the best fit.

\vspace{0.2cm}
\noindent{\it LHC four top search}. \ \
As discussed in Section~\ref{sec:LHCtopbounds}, after the $Z'$ is produced at the LHC, it can decay into $t\bar t$ according to Eq.~(\ref{eq:Z'decay}) leading to a final state containing four top quarks. This serves as an additional search channel for the $Z'$ boson with mass above $\sim$ 350\,GeV.  Although present bounds on the four top prouction cross section, $\sigma_{4t}=16.9_{-11.4}^{+13.8}\,$fb~\cite{Sirunyan:2017roi}, are not yet strong enough to constrain the $Z'$ mass if the uncertainties on this measurement scale as $1/\sqrt{\mathcal{L}}$ then future bounds will become significant.  In Fig.~\ref{fig:combined}, we show estimates for future bounds after 300 fb$^{-1}$ and 3000 fb$^{-1}$.

\vspace{0.2cm}
\noindent{\it Anomalous magnetic dipole of the muon}. \ \ 
It is intriguing to ask if the anomaly in the anomalous magnetic moment of the muon ($\Delta a_\mu$), which may be as large as $4.1\sigma$ \cite{Jegerlehner:2017lbd}, can be explained at the same time as explaining the $B$ anomalies.    The correction to the muon's magnetic moment coming from the $Z'$ is \cite{Queiroz:2014zfa},
\bea
\Delta a_\mu(Z') &=& \frac{z_\mu}{32\pi^2}\left[ (c_L+c_E)^2 \int_0^1 \frac{2x^2(1-x)}{(1-x)(1-z_\mu x) + z_\mu x}+ (c_L-c_E)^2 \int_0^1\frac{2x(1-x)(x-4)-4z_\mu x^3}{(1-x)(1-z_\mu x)+ z_\mu x}\right] \ , \nonumber \\
\eea
where $z_\mu=m_\mu^2/M_{Z'}^2$.
Unfortunately, we find no parameter space where both $\Delta a_\mu(Z')$ and the $B$ anomalies can be explained simultaneously.
On the other hand, the $\Delta a_\mu$ measurement also does not exclude the parameter space that could explain the $B$ anomalies.

\subsection{Interplay of all constraints in the effective $Z'$ model}

\begin{figure}[t]
   \centering    \includegraphics[width=0.9\textwidth]{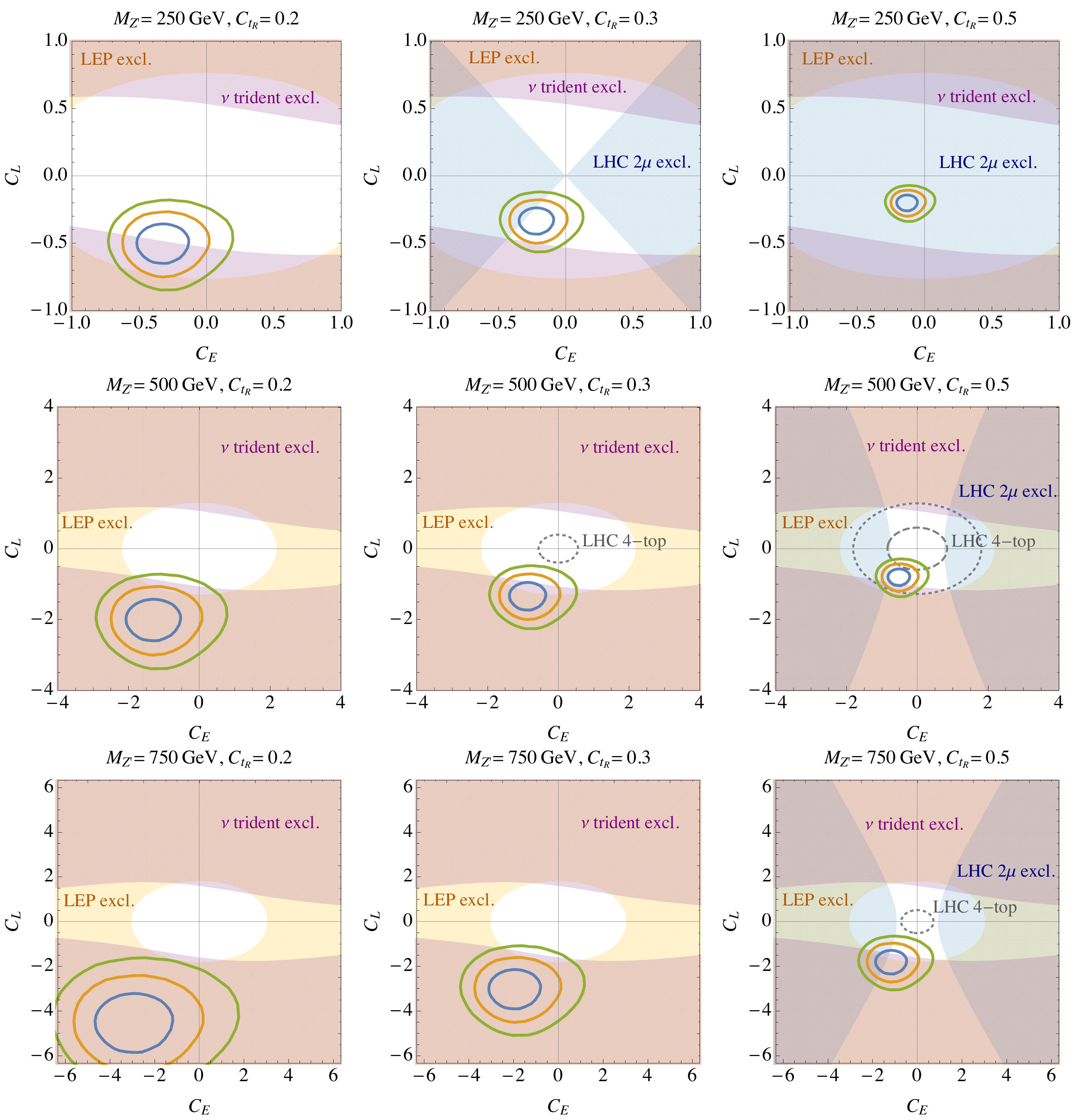}
   \caption{Favored regions and constraints on the parameter space of the effective $Z'$ model, defined in Eq.~(\ref{eq:IRLagrangian}). We fix $M_T=2\,$TeV in the calculations that involve the heavy vectorlike fermion $T$.  
The $1,2,3\,\sigma$ regions favored by $B$-physics anomalies in $R_K, R_K^*$ are enclosed by thick blue, orange and green contours, respectively.
The blue shaded region is excluded by the LHC dimuon resonance search, the magenta shaded region is excluded by the neutrino trident measurement, 
and the yellow shaded region is excluded by the LEP measurement of lepton flavor universality in $Z$-boson couplings.
The regions enclosed by the gray dashed (dotted) contours could potentially be excluded using the four top channel at LHC after 300 (3000)\, fb$^{-1}$, respectively.}
   \label{fig:combined}
\end{figure}

There are four parameters in the low energy model, Eq.~(\ref{eq:IRLagrangian}), $c_{t_R}$, $c_L$, $c_E$ and $m_{Z'}$.
In order to have a global view of the dependence on all these parameters, we combine all the constraints that are discussed above and show them in a series of plots
in Fig.~\ref{fig:combined}. We fix $M_{Z'}$ for each row of plots and fix $c_{t_R}$ for each column. In each plot, we show the constraints in the $c_L$-$c_E$ plane.
The $1,2,3\,\sigma$ favored regions by $B$-physics anomalies in $R_K, R_K^*$ are enclosed by thick blue, orange and green circles.
The blue shaded region is excluded by the LHC dimuon resonance search, the magenta shaded region is excluded by neutrino trident measurement, 
and the yellow shaded region is excluded by the LEP measurement of lepton flavor universality in $Z$-boson couplings. 
The uncolored shaded region is allowed when all these constraints are taken into account.
In addition, we find that the current four top search at LHC is not strong enough to place a bound in these plots. However, with a higher LHC luminosity, there will be a relevant bound for the case with $M_{Z'}\gtrsim 2m_t$ and $c_{t_R}\gtrsim0.3$. The regions enclosed by the gray dashed (dotted) circles could potentially be excluded with 300 (3000)\, fb$^{-1}$.

\section{Conclusion}\label{sec:conclusions}

In this work we considered the phenomenology of a $Z'$ boson coupling primarily to the (right-handed) SM top quark and leptons. Such a ``top-philic" $Z'$ could potentially have a mass around the weak-scale without contradicting current experimental constraints.  We first discussed the low-energy effective theory, which can be thought of as descending from a $U(1)'$ gauge theory, and presented two possible UV completions giving rise to the preferential coupling of the $Z'$ to the SM top quark.

At the LHC the top-philic $Z'$ can be produced in association with two top quarks at the tree-level or with a jet at the one-loop level. The $Z'+j$ channel, in particular, involves evaluating triangle diagrams containing one $Z'$ boson and two gluons,  which need to be dealt with carefully  in the presence of an anomalous $Z'$ current. The intricate interplay between the anomaly-cancelling spectator fermions in the UV and the low-energy phenomenology is examined. Furthermore, we present three recipes explaining how to properly compute the production cross-section of the $Z'+j$ channel in the effective theory.  When these subtleties are taken into account properly, in certain parts of parameter space, the production rate in the one-loop induced $Z'+j$ channel can be comparable to the rate in the $ttZ'$ channel. This corrects some mistakes in the earlier literature.

The top-philic $Z'$ can be looked for in a model-independent fashion at the LHC in the multi-top final states, which should be the focus of future searches. In addition to direct production of the $Z'$ it may be produced in decays of $T$, new QCD charged quarks predicted by the UV completion of the $U(1)'$ theory and expected to be not substantially heavier than the $Z'$.  Since no dedicated analyses are presently available, we estimated the bounds by recasting existing searches.  The $Z'$ could be as light as a few hundred GeV, while the $T$ must be heavier than $\sim 1.3$ TeV if all couplings of the $Z'$ other than to the top quark are turned off.

In addition to the multi-top final states, there are constraints from other channels, such as the inclusive dilepton resonance searches,  as well as probes of new physics in low-energy experiments. These low-energy probes include the $\nu$-trident production, precision electroweak constraints on $Z$-$Z'$ mixing, lepton universality measurements of the $Z$ boson at LEP, and muon $g-2$. A comprehensive study on the viable parameter space is presented. We apply these constraints to recent attempts to explain the experimental anomalies in $b\to s\mu^+\mu^-$ transitions. We found that part of the parameter space favored by the current $b\to s\mu\mu$ anomaly (in $R_K$, $R_{K^*}$ observables, etc) is already excluded by constraints from $\nu$ trident production. The future high-luminosity LHC will be able to cover much of the remaining regions capable of explaining this particular $B$ physics anomaly.

One of the purposes of this work is to point out the need to re-evaluate our search strategies for new physics, at this particular stage of the experimental program of the LHC, by demonstrating much remains to be done even for such a simple and well-studied extension of the SM like the $Z'$ boson. In particular, final states containing heavy flavor quarks such as the top quark have yet to be explored to their fullest potential. This is a new frontier waiting to be explored.
\\

\textbf{Note added:} While this work was being completed we became aware of~\cite{2016PoaU} where issues relating anomalies to $Z'$ production cross sections were discussed.

\section*{Acknowledgement}
We would like to acknowledge helpful discussions with Bill Bardeen, Bogdan Dobrescu, Chris Hill, Jernej Kamenik, Maxim Pospelov, Amarjit Soni, Yotam Soreq, and Jure Zupan.  PJF would especially like to thank David Tucker-Smith for earlier collaboration.
We thank Thomas Hahn for the help on {\tt FormCalc}.
This work is supported by the U.S. DOE grants DE-SC0010143 at Northwestern and  DE-AC02-06CH11357 at ANL.  This work was supported by the DoE under contract number DE-SC0007859 and Fermilab, operated by Fermi Research Alliance, LLC under contract number DE-AC02-07CH11359 with the United States Department of Energy.


\appendix

\section{Deriving the effective $Z'gg$ vertex via path integral}\label{pathintegral}

In this appendix, we formally derive the form factor Eq.~(\ref{Rosenberg}) in the heavy fermion limit from the path integral. 
As explained in Sections~\ref{sec:lowenergyEFT} and \ref{sec:Z'prod}, the three-point $Z'gg$ vertex is only calculable when the full theory is anomaly free.
Therefore, we include both $t$ and $T$ quarks in the discussion with their couplings to the $Z'$ boson related to each other (see Eq.~(\ref{anomalyfreecondition})), 
\begin{eqnarray}\label{eq:A1}
c_{t_L} - c_{t_R} =-( c_{T_L} - c_{T_R}) \ .
\end{eqnarray}

We start from the Lagrangian
\begin{eqnarray}
\mathcal{L} = \sum_{f=t, T} \bar f (i \slashed\partial - g_s \slashed G - c_{f_L} \slashed Z' P_L - c_{f_R} \slashed Z' P_R  - m_f) f \ ,
\end{eqnarray}
where $P_L=(1-\gamma_5)/2$, $G=G^a_\mu T^a$ is the gluon field and $T^a$ is the group generator. 
We treat the gauge fields $G$ and $Z'$ as space-time dependent background fields.

Because the $Z'$ coupling is chiral, the fermion mass $m_f$ is not invariant with respect to the corresponding $U(1)'$ gauge symmetry.
To restore the symmetry, one could promote $m_f$ into the VEV of a scalar field which is also charged under the $U(1)'$.
The path integral over the would-be goldstone modes in the scalar generates the Wess-Zumino term~\cite{DHoker:1984izu, DHoker:1984mif} which does not decouple in the large $m_f$ limit, and is proportional to $(c_{t_L} - c_{t_R}) + (c_{T_L} - c_{T_R})$.
The relation (\ref{eq:A1}) dictates that this term must vanish.

In addition, when $f=t, T$ is integrated out, there are also anomalous interacting terms between $Z'$ and $G$ at low energy which decouples in the large $m_f$ limit.
We derive these terms in the following.
Because the mass term $m_f$ is invariant under the $SU(3)_C$ gauge symmetry, the resulting operator will be automatically respect $SU(3)_C$.

For simplicity, we first set $c_{t_R}=c_{T_R}=0$, and keep only the left-handed couplings, $c_{t_L}, c_{T_L}$. 
The low energy effective action can be written as
\be
\Gamma[Z', G] = \log \int [\mathcal{D} f][\mathcal{D}\bar f] e^{i \int d^4x \mathcal{L}(x)}  
= \sum_{f=t, T} {\rm Tr}\log(i \slashed \partial - g_s \slashed G - c_{f_L} \slashed Z' P_L  - m_f) \ .
\ee

The 3-point $Z'gg$ vertex could be derived by Taylor expanding $\Gamma[Z', G]$ to the term of order $\mathcal{O}(g_s^2 c_{f_L})$, which, after some algebra, takes the form
\begin{eqnarray}
&&\hspace{-0.5cm}\frac{g_s^2 c_{f_L}}{3!} \left. \frac{\partial^3\Gamma[Z', G]}{\partial c_{f_L}\partial g_s \partial g_s} \right|_{g_s=g'=0} = - i 2 g_s^2 c_{f_L} {\rm tr}(T^a T^b) \int \frac{d^4\ell}{(2\pi)^4} \frac{1}{(\ell^2 - m_f^2)^3} 2 \left\{ (\partial\cdot Z') {\rm tr} \left[ (\slashed \partial \slashed G^a) (\slashed \partial \slashed G^b) P_L \rule{0mm}{4mm}\right] \rule{0mm}{4mm}\right. \nonumber \\
&& \hspace{5.8cm} \left. + {\rm tr} \left[ \left((\partial^2 \slashed G^a) (\slashed \partial \slashed G^b) + (\slashed \partial \slashed G^a) (\partial^2 \slashed G^b) \rule{0mm}{3.5mm}\right) \slashed Z' P_L \rule{0mm}{4mm}\right]
\rule{0mm}{4mm}\right\} \ .
\end{eqnarray}
It is straightforward to find the matrix element between the $\langle Z'_{\mu}(k_3) G^a_\sigma (k_1) G^b_\rho(k_2) |$ and $|0\rangle$ states, which gives
\begin{eqnarray}
- \sum_{f} \frac{i g_s^2 c_{f_L}}{48\pi^2 m_f^2}  {\rm tr}(T^a T^b) Z'^{\mu}_{k_3} G^{a\sigma}_{k_1} G^{b\rho}_{k_2}  \left\{ \varepsilon_{\alpha\beta\rho\sigma} k_{3\mu} k_1^{\alpha} k_2^\beta + \varepsilon_{\alpha\mu\rho\sigma} k_1^2 k_2^\alpha - \varepsilon_{\alpha\mu\rho\sigma} k_2^2 k_1^\alpha
\right\} \ .
\end{eqnarray}
This is the result when $f$ has only left-handed current coupling to the $Z'$ boson. If we also turn on both left- and right-handed couplings $c_{f_L}$ and $c_{f_R}$, the result will be
\begin{eqnarray}
-  \sum_{f} \frac{i g_s^2 (c_{f_L} - c_{f_R})}{48\pi^2 m_f^2}  {\rm tr}(T^a T^b) Z'^{\mu}_{k_3} G^{a\sigma}_{k_1} G^{b\rho}_{k_2}  \left\{ \varepsilon_{\alpha\beta\rho\sigma} k_{3\mu} k_1^{\alpha} k_2^\beta + \varepsilon_{\alpha\mu\rho\sigma} k_1^2 k_2^\alpha - \varepsilon_{\alpha\mu\rho\sigma} k_2^2 k_1^\alpha
\right\} \ ,
\end{eqnarray}
which is exactly the same form factor in the gauge invariant anomaly case at the large $m_f$ limit
(see Eqs.~(\ref{eq:Effvertex}) and (\ref{Rosenberg})).

In reality, because the top quark is not infinitely heavy at LHC energies, one must use the full form factor Eq.~(\ref{Rosenberg}) to reproduce the correct result for the top quark's contribution to the $Z'gg$ vertex.

\section{$b\to s\mu\mu$ in the gauged top model}\label{app:gaugedtop}

The gauged top model is a two-Higgs doublet model, with one Higgs $H_2$ charged under the new $U(1)'$ and the other $H_1$ uncharged. Because in the model $t_R$ and $L_\mu$ are also equally charged under the $U(1)'$, the VEV of $H_2$ is used to give the Dirac mass to the top quark and the muon. The VEV of $H_1$ is responsible for generating masses for other fermions. This setup does not lead to tree-level flavor change current (FCNC) mediated by the neutral Higgs bosons.

After electroweak symmetry breaking, we define the VEVs and the excitations of $H_{1,2}$ fields in the unitary gauge to be,
\be
\begin{split}
&H_1 = 
\begin{pmatrix}
-\sin\beta H^+ \\
\frac{1}{\sqrt2}(v_H\cos\beta + H_1 - i\sin\beta A)
\end{pmatrix} \ , \\
&H_2 = 
\begin{pmatrix}
\cos\beta H^+ \\
\frac{1}{\sqrt2}(v_H\sin\beta + H_2 + i\cos\beta A)
\end{pmatrix} \ ,
\end{split}
\ee
where $H^\pm$ is the physical charged scalar, and the lighter of the neutral scalars (a linear combination of $H_{1,2}$) can be made SM-like if the model is close to the alignment limit. This could be realized in the decoupling limit of the second doublet, in which case $m_{H^\pm} \approx m_{H} \approx m_A \gg v_H =246\,$GeV.
Here for simplicity we assume the two Higgs doublet VEVs preserves CP.
In order for the model to be realistic, we also introduce a SM singlet scalar $\Phi$ which also carries $U(1)'$ charge unit $q_t$, with VEV $v_{\Phi'}/\sqrt{2}$.
Because $H_2$ is charged under $SU(2)_L$, $U(1)_Y$ and $U(1)'$, its VEV will induce a mixing among all the gauge bosons. In the basis of $(Z_{\rm SM}, Z')$ (where $Z_{\rm SM}$ is the $Z$ boson in the SM limit), the mass matrix takes the form
\be\label{eq:Mgauge}
M^2 = \begin{pmatrix}
\frac{1}{4\cos^2\theta_W} g_2^2 v_H^2 & \frac{1}{2\cos\theta_W} g_2 g_X v_H^2\sin^2\beta \\
\frac{1}{2\cos\theta_W} g_2 g_X v_H^2\sin^2\beta & g_X^2 (v_H^2\sin^2\beta + v_{\Phi'}^2) 
\end{pmatrix} \ ,
\ee
where $g_2$ is the gauge coupling for the SM $SU(2)_L$.
Without loss of generality, we set $q_t=1$ for simplicity. The LEP measurements dictate the off-diagonal element $M^2_{12}$, which controls the $Z_{\rm SM}-Z'$ mixing, to be small compared to the diagonal elements. 

We have explicitly calculated the process $b\to s\mu\mu$ in this model, which occurs at one-loop level. It is useful to classify the various contributions diagrammatically into the following groups. We use the mass insertion method, treating $(M^2)_{12}/M_{Z'}^2$ (where $M_{Z'}^2 = (M^2)_{22} \simeq g_X^2 (v_H^2\sin^2\beta + v_{\Phi'}^2)$) as the small parameter. We work in the unitary gauge.
\begin{itemize}
\item The first class of diagrams correspond to the SM contribution. They include the $Z, \gamma$ penguin diagrams as well as the box diagram with $W^\pm$ boson exchange. These contributions are of order $\mathcal{O}((M^2)_{12}/M_{Z'}^2)^0$. We do not write down their explicit forms, but refer to the classic paper~\cite{Inami:1980fz}.

Instead, we write down the $Z_{\rm SM}$ penguin vertex which will be useful later~\cite{He:2009rz},
\begin{figure}[h]
\includegraphics[width=0.18\textwidth]{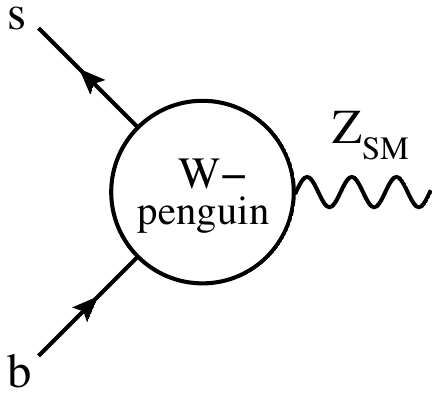}\raisebox{1.2cm}{\ $\displaystyle{=i\frac{m_t^2 V_{tb} V^*_{ts}}{256\pi^2 M_W^2} \left[ 2 \left( \frac{1}{\epsilon} + \ln \frac{\mu^2}{m_t^2} \right) - \frac{m_t^2 - 7 M_W^2}{m_t^2 - M_W^2}
- \frac{6 M_W^4}{(m_t^2 - M_W^2)^2} \ln\frac{m_t^2}{M_W^2} \right]} \bar s_L \gamma^\mu b_L  Z_{\rm SM,\mu} \ ,$}\vspace{-0.3cm}
\end{figure}

and we define the ``W-$Z_{\rm SM}$ penguin'' to be the sum of diagrams with the $Z_{\rm SM}$ gauge boson attached in all possible ways,
\begin{figure}[h]
\centering\includegraphics[width=1\textwidth]{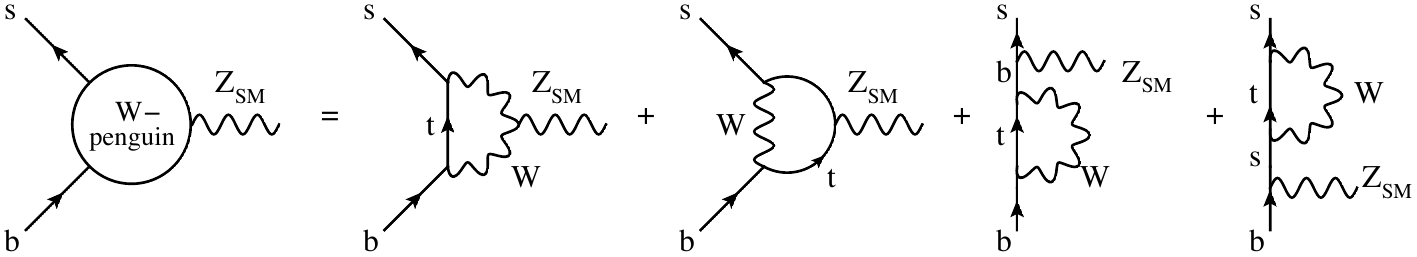}
\end{figure}

\item The second class of diagrams corresponds to dressing the $W$-$Z_{\rm SM}$ (as well as the $W$-$Z'$) penguin diagrams with all possible $Z_{\rm SM}-Z'$ mixing mass insertions (blue cross in the diagrams below), that are up to order $\mathcal{O}((M^2)_{12}/M_{Z'}^2)^1$,
\begin{figure}[h]
\centering\includegraphics[width=1\textwidth]{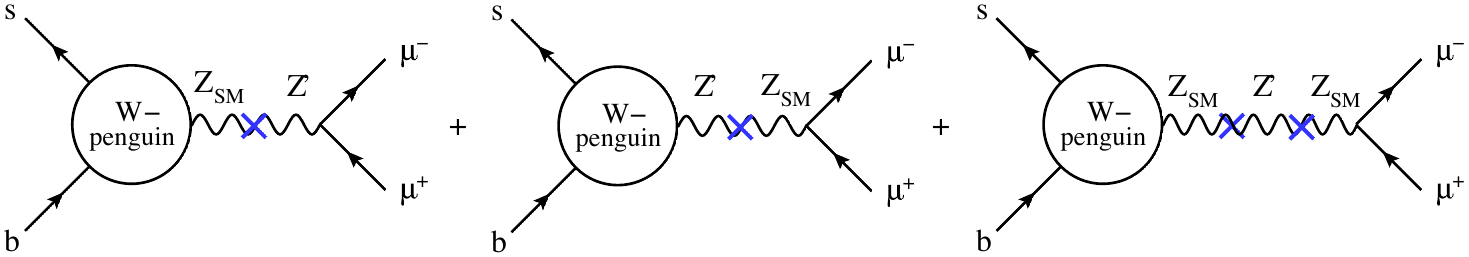}\vspace{-0.3cm}
\end{figure}

Their contributions to the effective operator $(\bar s_L \gamma^\mu b_L)(\bar \mu_L \gamma_\mu \mu_L)$ take the form
\be\label{eq:part1}
\begin{split}
\mathcal{A}_{W\text{-penguin-}Z'}&= - i\frac{g^2 g_X^2 m_t^2 V_{tb} V^*_{ts}\cos^2\beta (1-\sin^2\beta \cos2\theta_W)}{128\pi^2 M_{Z'}^2 M_W^2} (\bar s_L \gamma^\mu b_L)(\bar \mu_L \gamma_\mu \mu_L) \\ 
&\times\left[ 2 \left( \frac{1}{\epsilon} + \ln \frac{\mu^2}{m_t^2} \right) - \frac{m_t^2 - 7 M_W^2}{m_t^2 - M_W^2}
- \frac{6 M_W^4}{(m_t^2 - M_W^2)^2} \ln\frac{m_t^2}{M_W^2} \right] \ ,
\end{split}
\ee
which is infinite, but is proportional to $\cos^2\beta$. We find a key relation between the $W$-$Z_{\rm SM}$ and the $W$-$Z'$ penguin vertices which is crucial for deriving the above result,
\begin{figure}[h]
\centering\includegraphics[width=0.18\textwidth]{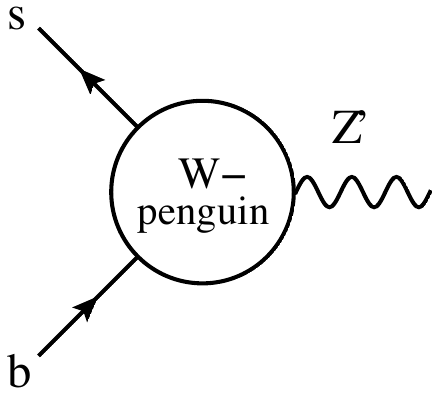}\raisebox{1.2cm}{\ $\displaystyle{=}$}
\includegraphics[width=0.18\textwidth]{W-penguin.pdf}
\raisebox{1.2cm}{\ $\displaystyle{\times \left( -\frac{2g_X\cos\theta_W}{g} \right) \times \frac{\bar s_L \gamma^\mu b_L  Z'_\mu}{\bar s_L \gamma^\mu b_L  Z_{\rm SM,\mu}}} \ .$}
\vspace{-0.3cm}
\end{figure}

\item The third class of diagrams are those that exist in the normal two-Higgs-doublet model (2HDM), with charged-Higgs penguin diagrams and $Z_{\rm SM}, \gamma$ exchange, as shown below,
\begin{figure}[h]
\centering\includegraphics[width=0.33\textwidth]{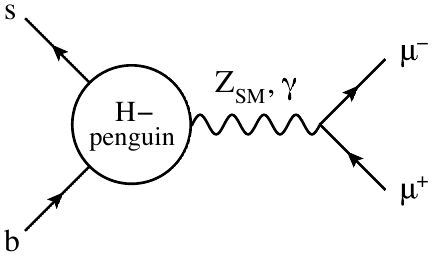}\vspace{-0.3cm}
\end{figure}

where the ``H-penguin'' vertex is defined similarly as the ``W-penguin'' diagrams.
They are the only new contributions to $b\to s\mu\mu$ process in normal 2HDM, and the $Z_{\rm SM}$ and $\gamma$ exchange diagrams are finite individually.
In the heavy charged-Higgs limit, these contributions must go as $1/M_{H^\pm}^2$. Because we are more interested in the contributions from $Z'$ exchange, in this paper, we decide to work with heavy enough $H^\pm$ ($M_{H^\pm}=10\,$TeV) so that these contributions are negligible.
Similarly, the box diagrams involving the charged Higgs are also suppressed by its large mass, as well as by the small lepton Yukawa couplings.

\item The fourth class of diagrams correspond to dressing the above $H^\pm$-$Z_{\rm SM}$ penguin diagrams with  $Z_{\rm SM}-Z'$ mixing mass insertions, up to order $\mathcal{O}((M^2)_{12}/M_{Z'}^2)^1$. The diagrams are shown in below,
\begin{figure}[h]
\centering\includegraphics[width=0.75\textwidth]{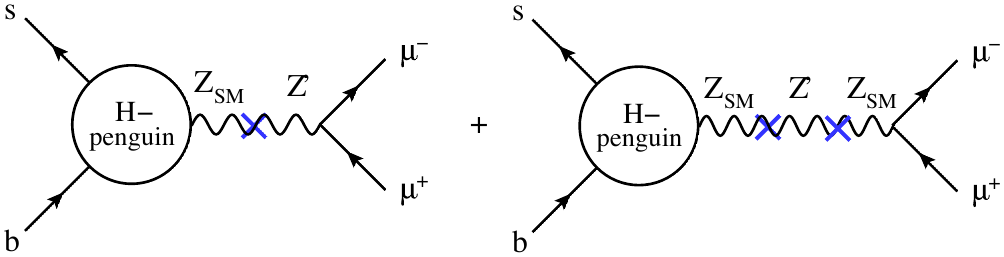}\vspace{-0.3cm}
\end{figure}

Because the $H^\pm$-$Z_{\rm SM}$ vertex is finite, these diagrams are also finite, and go as $1/M_{H^\pm}^2$ in the heavy $H^\pm$ limit. Similar to bullet 3, we will take a large value of $M_{H^\pm}=10\,$TeV and neglect these contributions.

\item The fifth class of diagrams correspond to dressing the $H^\pm$-$Z'$ penguin diagrams (similar to the above $H^\pm$-$Z_{\rm SM}$ penguin but with $Z_{\rm SM}$ replace by $Z'$ gauge boson) with possible $Z_{\rm SM}-Z'$ mixing mass insertions, up to order $\mathcal{O}((M^2)_{12}/M_{Z'}^2)^1$. The diagrams are shown in below,
\begin{figure}[h]
\centering\includegraphics[width=0.72\textwidth]{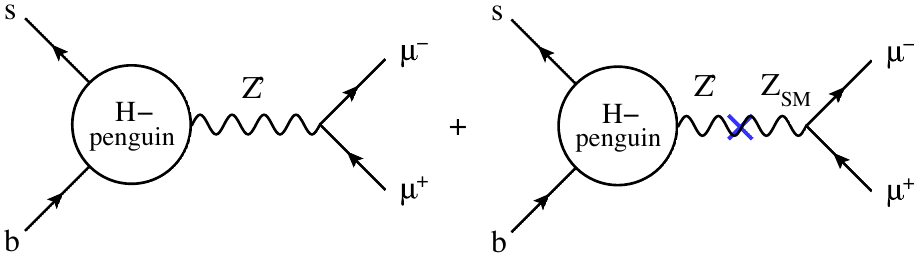}\vspace{-0.3cm}
\end{figure}

We find these diagrams are infinite and their contributions to the effective operator $(\bar s_L \gamma^\mu b_L)(\bar \mu_L \gamma_\mu \mu_L)$ take the form
\be\label{eq:part2}
\begin{split}
\mathcal{A}_{H\text{-penguin-}Z'}&= i\frac{g^2 g_X^2 m_t^2 V_{tb} V^*_{ts}\cos^2\beta (1-\sin^2\beta \cos2\theta_W)}{128\pi^2 M_{Z'}^2 M_W^2} (\bar s_L \gamma^\mu b_L)(\bar \mu_L \gamma_\mu \mu_L) \\
&\times  \left[ 2 \left( \frac{1}{\epsilon} + \ln \frac{\mu^2}{M_{H^\pm}^2} \right) +1 + \mathcal{O}\left(\frac{m_t^2}{M_{H^\pm}^2}\right) \right] \ ,
\end{split}
\ee
whose infinite part exactly cancels with that in Eq.~(\ref{eq:part1}).

\end{itemize}

\noindent To summarize, we find that in the heavy $H^\pm$ limit, only bullets 2 and 5 in the above make non-zero contribution to the $(\bar s_L \gamma^\mu b_L)(\bar \mu_L \gamma_\mu \mu_L)$ operator. Summing Eqs.~(\ref{eq:part1}) and (\ref{eq:part2}) together and compare them with the standard definition of $O_{9,10}$ operators, we derive
\be
\begin{split}
\delta C_{9} = & -\frac{g_X^2 m_t^2 \cos^2\beta (1-\sin^2\beta \cos2\theta_W)}{4 M_{Z'}^2 e^2} 
\left[ \ln \frac{M_{H^\pm}^2}{m_t^2} - \frac{m_t^2 - 4 M_W^2}{m_t^2 - M_W^2} - \frac{3M_W^4}{(m_t^2 - M_W^2)^2}\ln\frac{m_t^2}{M_W^2} \right]+\mathcal{O}\left( \frac{m_t^2}{M_{H^\pm}^2} \right) \\
=& - C_{10}\ .
\end{split} 
\ee

\bibliography{ZpUV}
\bibliographystyle{JHEP}

\end{document}